\documentclass[twocolumn,english,aps,prb,showpacs,byrevtex,amsmath,amssymb,superscriptaddress]{revtex4-1}

\usepackage{graphicx}
\usepackage{epstopdf}
\usepackage{dcolumn}
\usepackage{bm}
\usepackage{gensymb}
\usepackage{upgreek}
\usepackage{hyperref}
\usepackage{graphicx}
\usepackage{epstopdf}
\usepackage{dcolumn}
\usepackage{bm}
\usepackage{gensymb}
\usepackage{upgreek}
\usepackage{booktabs}
\usepackage{hyperref}

\usepackage{amssymb,mathtools}
\usepackage{color}
\usepackage{amsmath} 

\usepackage{tikz,xcolor,hyperref}
\definecolor{lime}{HTML}{A6CE39}
\DeclareRobustCommand{\orcidicon}{%
	\begin{tikzpicture}
	\draw[lime, fill=lime] (0,0)
	circle [radius=0.16]
	node[white] {{\fontfamily{qag}\selectfont \tiny ID}};
	\draw[white, fill=white] (-0.0625,0.095)
	circle [radius=0.007];
	\end{tikzpicture}
	\hspace{-2mm}
}

\foreach \x in {A, ..., Z}{%
	\expandafter\xdef\csname orcid\x\endcsname{\noexpand\href{https://orcid.org/\csname orcidauthor\x\endcsname}{\noexpand\orcidicon}}
}

\begin{document}

\title{Spin-orbit insulating phase in SnTe cubic nanowires: \\ consequences on the topological surface states}

\author{Ghulam Hussain\orcidE}
\email{ghussain@magtop.ifpan.edu.pl}
\affiliation{International Research Centre MagTop, Institute of Physics, Polish Academy of Sciences, Aleja Lotnik\'ow 32/46, PL-02668 Warsaw, Poland}

\author{Kinga Warda}
\affiliation{International Research Centre MagTop, Institute of Physics, Polish Academy of Sciences, Aleja Lotnik\'ow 32/46, PL-02668 Warsaw, Poland}
\affiliation{Faculty of Applied Physics and Mathematics, Gdansk University of Technology, Gda\'nsk 80-233, Poland}

\author{Giuseppe Cuono\orcidD}
\email{gcuono@magtop.ifpan.edu.pl}
\affiliation{International Research Centre MagTop, Institute of Physics, Polish Academy of Sciences, Aleja Lotnik\'ow 32/46, PL-02668 Warsaw, Poland}

\author{Carmine Autieri\orcidA}
\email{autieri@magtop.ifpan.edu.pl}
\affiliation{International Research Centre MagTop, Institute of Physics, Polish Academy of Sciences, Aleja Lotnik\'ow 32/46, PL-02668 Warsaw, Poland}

\begin{abstract}
We investigate the electronic, structural and topological properties of the SnTe and PbTe cubic nanowires using \textit{ab initio} calculations. Using standard and linear-scale density functional theory, we go from the ultrathin limit up to the nanowires thicknesses observed experimentally. Finite-size effects in the ultra-thin limit produce an electric quadrupole and associated structural distortions, these distortions increase the band gap but they get reduced with the size of the nanowires and become less and less relevant. Ultrathin SnTe cubic nanowires are trivial band gap insulators, we demonstrate that by increasing the thickness there is an electronic transition to a spin-orbit insulating phase due to trivial surface states in the regime of thin nanowires. These trivial surface states with a spin-orbit gap of a few meV appear at the same $k$-point of the topological surface states.
Going to the limit of thick nanowires, we should observe the transition to the topological crystalline insulating phase with the presence of two massive surface Dirac fermions hybridized with the persisting trivial surface states. Therefore, we have the co-presence of massive Dirac surface states and trivial surface states close to the Fermi level in the same region of the $k$-space.
According to our estimation, the cubic SnTe nanowires are trivial insulators below the critical thickness t$_{c1}$=10 nm, and they become spin-orbit insulators between t$_{c1}$=10 nm and t$_{c2}$=17 nm, while they transit to the topological phase above the critical thickness of t$_{c2}$=17 nm. These critical thickness values are in the range of the typical experimental thicknesses, making the thickness a relevant parameter for the synthesis of topological cubic nanowires. Pb$_{1-x}$Sn$_x$Te nanowires would have both these critical thicknesses t$_{c1}$ and t$_{c2}$ at larger values depending on the doping concentration.
\end{abstract}

\date{\today}
\maketitle

\section{Introduction}

Majorana fermions are intriguing particles that have garnered significant attention in the field of condensed matter physics. Majorana fermions are non-Abelian anyons, meaning their quantum states exhibit intriguing noncommutative properties. They obey non-Abelian statistics and their topological protection could be crucial for quantum computing and topological quantum information processing. Their potential application in topological quantum computation promises quantum algorithms that are robust against decoherence and errors. Majorana-based qubits are expected to be more robust than other qubit technologies due to their topological protection and are envisioned to play a vital role in quantum error correction protocols\cite{Sau2020,Beenakker2013}.

One of the platforms proposed to host Majorana fermions is a topological nanowire with superconductivity in an applied magnetic field.
However, the realization of Majorana's fermions is challenging as it requires the growth of a nanowire crystal to cultivate a column of atoms measuring approximately 100 nm in diameter. Subsequently, this system needs to be integrated with a circuit that exhibits the necessary sensitivity to monitor individual electrons as they traverse through it. Moreover, all of this intricate work must be conducted at temperatures merely one-hundredth of a degree above absolute zero and within a magnetic field that is 10,000 times stronger than Earth’s \cite{Frolov21}.
Nanowires made from various materials like semiconductor-superconductor hybrids or topological insulators with proximity-induced superconductivity have been studied to realize Majorana fermions\cite{Stanescu2013}.

IV-VI semiconductors have exhibited interesting properties such as thermoelectricity \cite{wang2011heavily, wood1988materials}, ferroelectricity\cite{lebedev1994ferroelectric, liu2020synthesis} and superconductivity\cite{matsushita2006type, mazur2019experimental}. In particular, the discovery of topological crystalline insulating phase (TCI) in SnTe\cite{fu2011topological, hsieh2012topological,Lau19,Tanaka07,Okada13} and some of their substitutional alloys such as Pb$_{1-x}$Sn$_x$Te\cite{xu2012observation}
and Pb$_{1-x}$Sn$_x$Se\cite{dziawa2012topological,Kazakov21} have aroused tremendous research interest to further study this class of materials. 
The TCI phase is different from the conventional topological insulating phase, indeed the linearly dispersing Dirac states on the high-symmetry surfaces of the TCIs are protected by crystal symmetries \cite{Fu07} and not by time-reversal symmetry.
SnTe and its substitutional alloys are TCIs with mirror symmetry, and the presence of the surface states is indicated by a non-zero integer topological invariant named mirror Chern number \cite{Teo08,Safaei_2015,hsieh2012topological,Cuono22,Rechcinski21}.
Furthermore, it has been shown that SnTe is a helical higher-order topological insulator \cite{Schindler18,Kooi20,vanMiert18}.
It has been observed that the characteristic properties change by changing the size or dimensions (2D or 3D phase) of materials, namely the properties will change if we move from bulk \cite{Plekhanov14,Wang20,Barone13} to thin films \cite{Slawinska20,Volobuev17,Liu18}. For instance, SnTe is a trivial insulator at low thickness but becomes topological above some critical thickness\cite{liu2014spin, liu2015crystal} quite robust against impurity doping\cite{D1NR07120C}. 
Recently, the twinning in thin films as a function of the mirror Chern number was studied\cite{Samadi23}.
By using scanning tunneling microscopy for Pb$_{1-x}$Sn$_x$Se, it was shown that atomically flat terraces in the Se sublattice separated by step edges of various heights are present. Enhancements of the local DOS have been found at odd step edges, and by using tight binding models spin-polarized flat bands connecting Dirac points have been shown \cite{Sessi16}. The properties of the low-energy states at the surface atomic steps and the behavior of such edge channels under doping were investigated \cite{Sessi16,Brzezicki19,mazur2019experimental,Wagner12}.

Going to the one-dimensional (1D) case, the topological crystalline insulator SnTe nanowires (NWs) can be even more interesting than the bulk due to a larger contribution from the topological surface states and to the 1D confinement effect\cite{C4NR05124F}. Methods to produce SnTe NWs by using graphene\cite{Sadowski2018}, high yield and alloy nanoparticles as growth catalysts\cite{liu2020synthesis} or to obtain the smallest possible SnTe NWs by using single-walled carbon nanotubes\cite{Vasylenko18} were found.
TCI NWs are a versatile platform for the confinement and manipulation of Dirac fermions \cite{Skiff22}. The Pb$_{1-x}$Sn$_{x}$Te cubic NWs were also synthesized \cite{Safdar15,dad2022nearly,Saghir15,Xu2016-vw}.
Recently, pentagonal nanowires with a perfect C$_5$ symmetry have been synthesized in Pb$_{1-x}$Sn$_{x}$Te. This non-cubic crystal structure (impossible in a pure ionic system) rises from the mixture of covalent and ionic bonds. Additionally, the pentagonal phase could be a platform for high-order topology. The results of the investigation on pentagonal nanowires will be presented elsewhere\cite{Hussain23pentagonal}.

Theoretically, the SnTe NWs were  investigated and it was shown that different topological states can be realized under the application of various magnetic fields \cite{Nguyen22}, bulk Majorana modes are present when there is inversion symmetry, while under symmetry-breaking fields at the ends of the NWs, Majorana zero modes appear while the Majorana bulk modes are gapped \cite{Nguyen22}.
Ultrathin SnS and SnSe NWs were also studied by means of \textit{ab initio} calculations \cite{Shukla22}, while the NWs of trivial PbTe were shown to be also interesting for the possible presence of Majorana zero modes\cite{Song23,Jung22}.

The 3D and 2D cases of these rock-salt chalcogenides are widely investigated from first principles, while the one-dimensional structures like NWs are poorly investigated so far due to the huge number of atoms to be considered when we break the boundary conditions along 2 dimensions. We find that the thin cubic NWs show a robust insulating behavior. 
We investigate the evolution of the band gap as a function of the thickness for the cubic NWs and we find the critical thickness at which the system becomes topological.
The paper is organized as follows: in the next Section the computational details are described, the third section is devoted to the theoretical investigation of the structural and the electronic properties of the cubic NWs, while in the fourth section, the topology of the NWs is investigated. In the last Section, we draw our conclusions.

\begin{figure}[t!]
\centering
\includegraphics[scale=0.08]{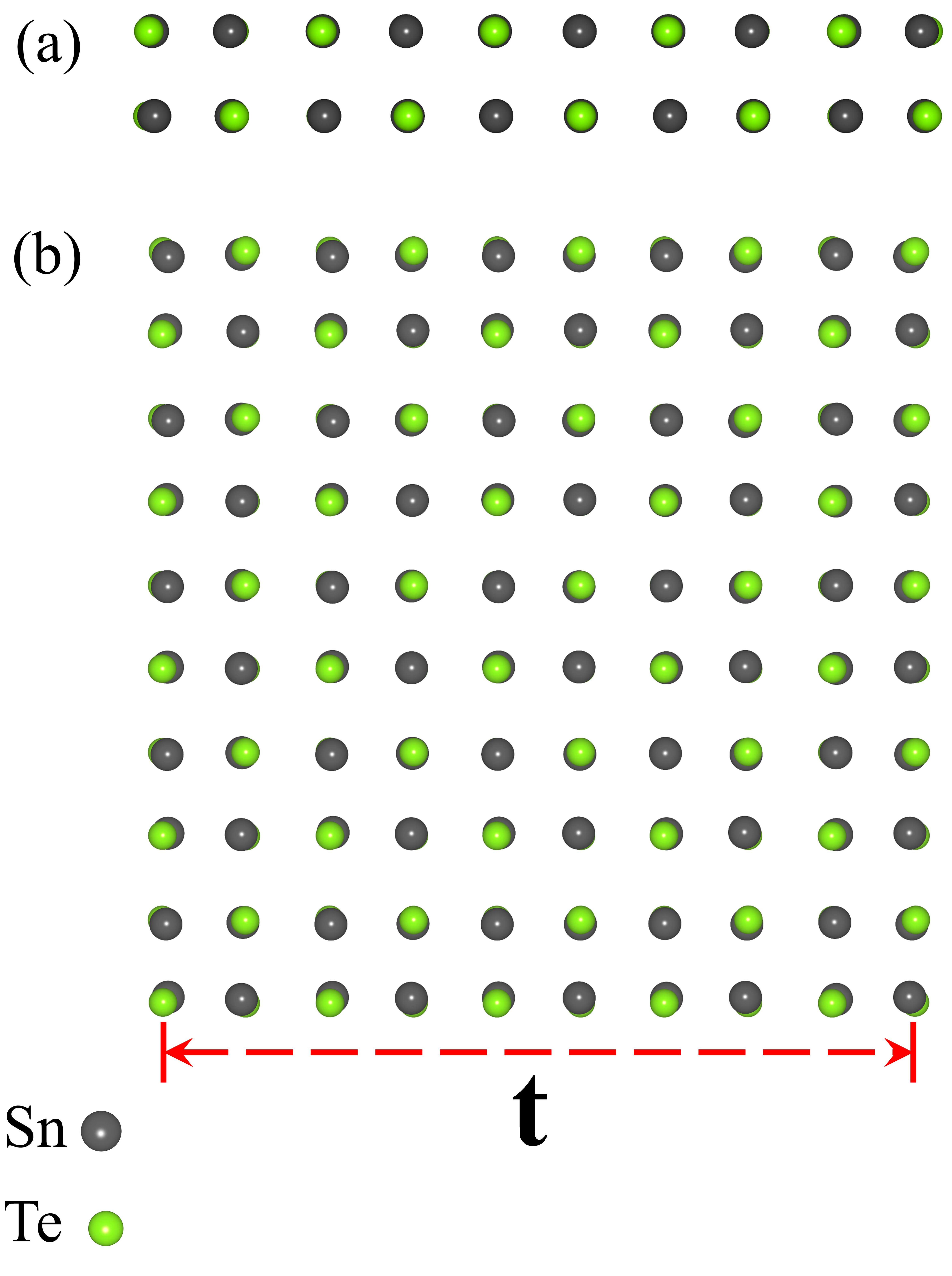}
\caption{a) Side view and b) top view of the crystal structure of the cubic SnTe nanowires with an even number of atoms along the x- and y-direction. We indicate '$t$' as the thickness of the nanowire. The unit cell is composed of 2N$\times$2N$\times$2 atoms with periodic boundary conditions along the z-axis.}
\label{crystal_structure_cubic}
\end{figure} 

\section{Computational details} 

The results were obtained within the framework of the first-principles density functional theory (DFT), the calculations are fully relativistic by considering spin-orbit coupling (SOC) if not mentioned otherwise. The system presents inversion symmetry since we do not assume ferroelectricity, as a consequence, the systems host the Kramer's degeneracy due to the combination of time-reversal and inversion symmetry. Therefore, all bands reported in this paper are double degenerate and we plot the band structure only along the $k$-path $\Gamma$-Z. The system has a mirror along the (110) and (1$\overline{1}$0) planes and a non-symmorphic symmetry respect to the (100) and (010) planes.

We investigated the cubic SnTe and PbTe NWs with [001] orientation.
The lattice constant $c$ is equal to 6.2964 and 6.4277 {\AA} for SnTe and PbTe, respectively, which are the experimental values.
These lattice constants are equivalent to twice the anion-cation distance. 
We limit the study to an even number of atoms along the x- and y-direction to avoid odd-even effects and to keep the crystal structure stoichiometric. Indeed, depending on the odd or even number of atoms the system hosts the mirror or non-symmorphic symmetry\cite{Brzezicki19}. For the VASP calculations, we investigate unit cells containing 2N$\times$2N$\times$2 nanowires, where N=1, 2, 3... is the number of atoms as we can see in Fig. \ref{crystal_structure_cubic}a). When we use ONETEP, we double the cell along the z-axis studying 2N$\times$2N$\times$4 nanowires with the lattice constant of the superlattice $c_{sup}$=2$c$. We define $t$ as the in-plane thicknesses of the nanowires as shown in Fig. \ref{crystal_structure_cubic}b).

Structural and electronic structure calculations of the thin nanowires were performed with a plane wave basis set and projector augmented wave method using the VASP\cite{VASP} package. A plane-wave energy cut-off of 250~eV has been used. As an exchange-correlation functional, the generalized gradient approximation (GGA) of Perdrew, Burke, and Ernzerhof (PBE) has been adopted\cite{perdew1996generalized}. 
The description of the IV-VI semiconductors and more in general narrow band semiconductors suffers of the band gap problem that overestimates the topological region of the phase diagram giving a wrong band ordering. To ease this problem and obtain a description closer to the experiments, we need to go beyond the standard GGA approximation \cite{Cuono22,Cuono23Eu,Islam19,Hussain22}.
In the case of nanowires, a more realistic band order and band gap can be obtained with the meta-GGA approach named SCAN, namely, the strongly constrained and appropriately normed functional \cite{PhysRevLett.115.036402}, which is one of the few exchange-correlation functional working well in the presence of SnTe and PbTe surfaces. Both results with GGA and meta-GGA are qualitatively similar but the meta-GGA gives a larger gap that is much closer to the experimental results. We have performed the VASP self-consistent calculations using 1$\times$1$\times$17 for GGA and 1$\times$1$\times$12 $k$-points centered in $\Gamma$ for SCAN, while for the band structures, we have used 60 $k$-points from $\Gamma$ to Z.

The number of atoms in the system is N$_{ATOMS}$=8N$^2$, standard DFT codes scale as O(N$_{ATOMS}^3$), therefore, they are scaling as the O(N$^6$). This poses serious limitations in the study of thick NWs.
In order to increase the thickness of cubic SnTe NWs, we need to study a large number of atoms with SOC. This was accomplished through the use of ONETEP \cite{10.1063/1.1839852,10.1063/5.0004445}, a linear-scaling density functional theory (LS-DFT) approach. In the ONETEP code orbitals that are nonorthogonal generalized Wannier functions (NGWFs) are expressed within the basis of periodic sinc (psinc) functions, providing accuracy comparable to the traditional plane-wave DFT software \cite{10.1063/1.2796168}. Simultaneously, the developed approach offers parallel computation, which allows the simulations of thousands of atoms \cite{Skylaris_2008}. The ONETEP calculations were performed with the use of PBE functional and PAW pseudopotentials from the JTH library \cite{JOLLET20141246}. The investigation of the bandstructure was carried out with the use of spectral function unfolding methodology \cite{PhysRevB.91.195416}. The sampling of the reciprocal space consisted of 110 $k$-points, spaced equally along the z-reciprocal lattice vector. With each of the Sn and Te atoms, there were assigned 9 NGWFs. Considering the study of large systems, for all NGWFs the radius was determined as $R_{\phi}\textnormal{ = 5 }a_{0}$, allowing the convergence and optimal accuracy to be maintained. To ensure the requirements of the bandstructure unfolding formalism, the diameter of the NGWFs could not exceed half the length of the simulation cell. Therefore, the size of the nanowire in the axial direction had to be doubled compared to the structure studied within VASP. The basis set consisted of the psinc functions distant from each other by $d_{\textnormal{psinc}}$ = 0.36 \AA, resulting in the density kernel cutoff $E_{\textnormal{c}}$ equal 447 eV. The length of the simulation cell in the z-direction was equal to $c_{sup}$ = 12.5928 \AA, with the imposition of periodic boundary conditions (PBCs). The lengths along the x- and y-directions were selected as multiplies of the $d_{\textnormal{psinc}}$ depending on the thickness of the structure, with the vacuum spacing applied. The system investigated consisted of up to 784 atoms for the SOC calculations and 2704 atoms for non-relativistic calculations. The density kernel truncation was not applied.
Considering that ONETEP is based on localized Wannier functions and it is developed for molecules and surfaces, the surface states calculated within ONETEP would be quite reliable.

\begin{figure}[t]
	\centering
\includegraphics[width=1.0\columnwidth, angle=0]{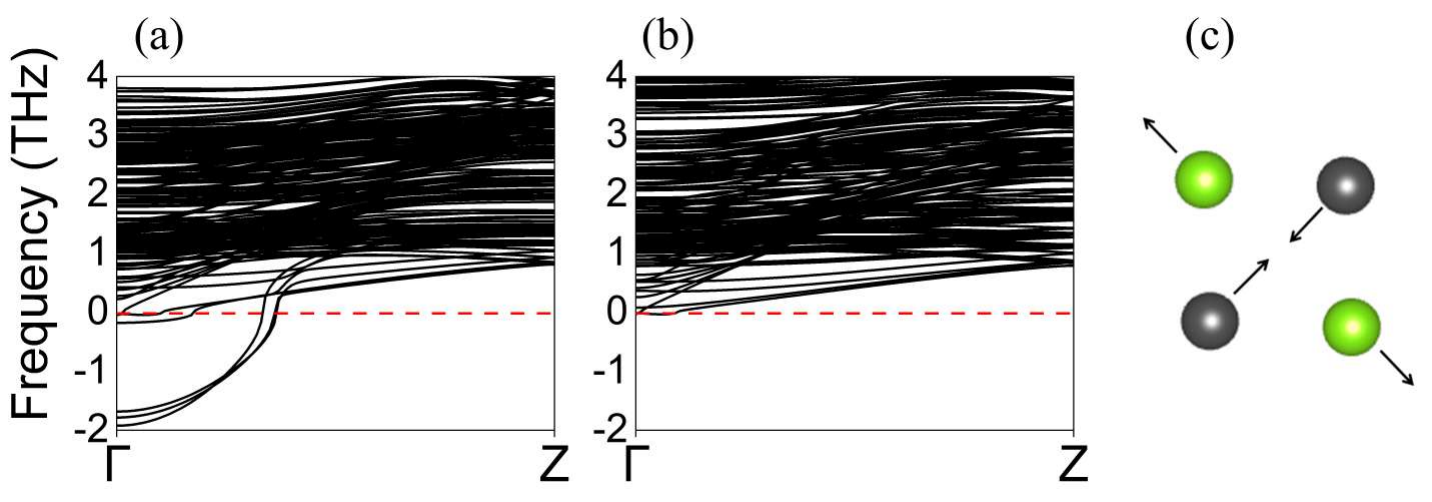}
\caption{Phonon dispersions for undistorted (a) and distorted (b) cubic NWs. This indicates that the distortions in the cubic systems are important for the dynamical
stability. (c) Quadrupolar distortion in the case of the 2$\times$2 SnTe nanowire. The arrows represent the displacements with respect to the ideal NaCl crystal structure. As in Fig. \ref{crystal_structure_cubic}, the green balls represent the Te atoms, while the dark grey balls represent the Sn atoms.}
\label{quadruplar}

\end{figure}
\begin{figure}[th]
\centering
\includegraphics[scale=0.33]{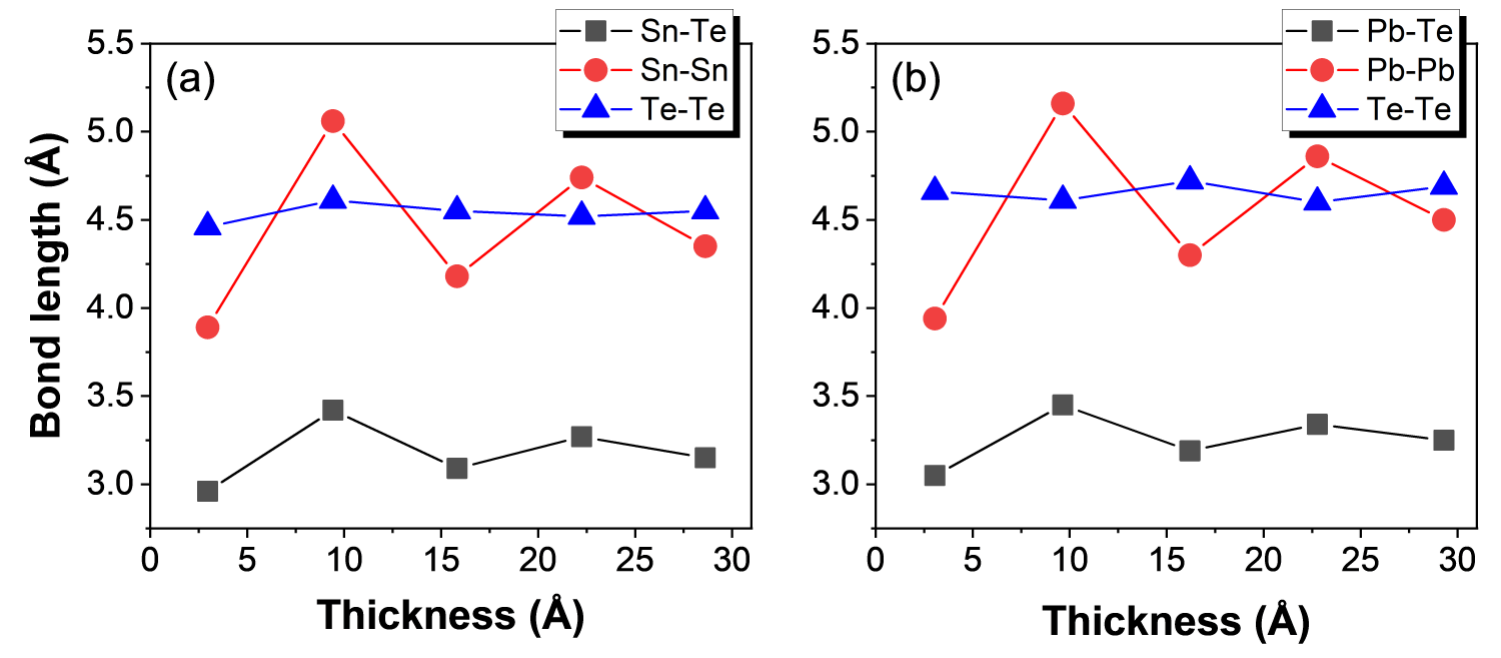}
\caption{Bond lengths for [001] oriented cubic nanowires as a function thickness for (a) SnTe and (b) PbTe cubic nanowires, respectively. The anion-cation distance is the first neighbor distance, while the anion-anion and cation-cation distances are the second neighbor distances. The second neighbor distances are larger by a factor $\sqrt{2}$ with respect to the first neighbor distances. The line is a guide to the data points.}
\label{Bond length}
\end{figure}

\section{Structural and Electronic properties of the S\lowercase{n}T\lowercase{e} and P\lowercase{b}T\lowercase{e} cubic nanowires}

In this Section, we investigate the structural and electronic properties of the NWs grown along the [001] direction focusing on the ultrathin nanowires using the VASP code. The (001) surfaces in cubic SnTe are not polar, the charge present on the inner atoms is almost the same of the surface atoms, therefore, we do not need to introduce dangling bonds.

\begin{figure}[th]
\centering
\includegraphics[scale=0.38]{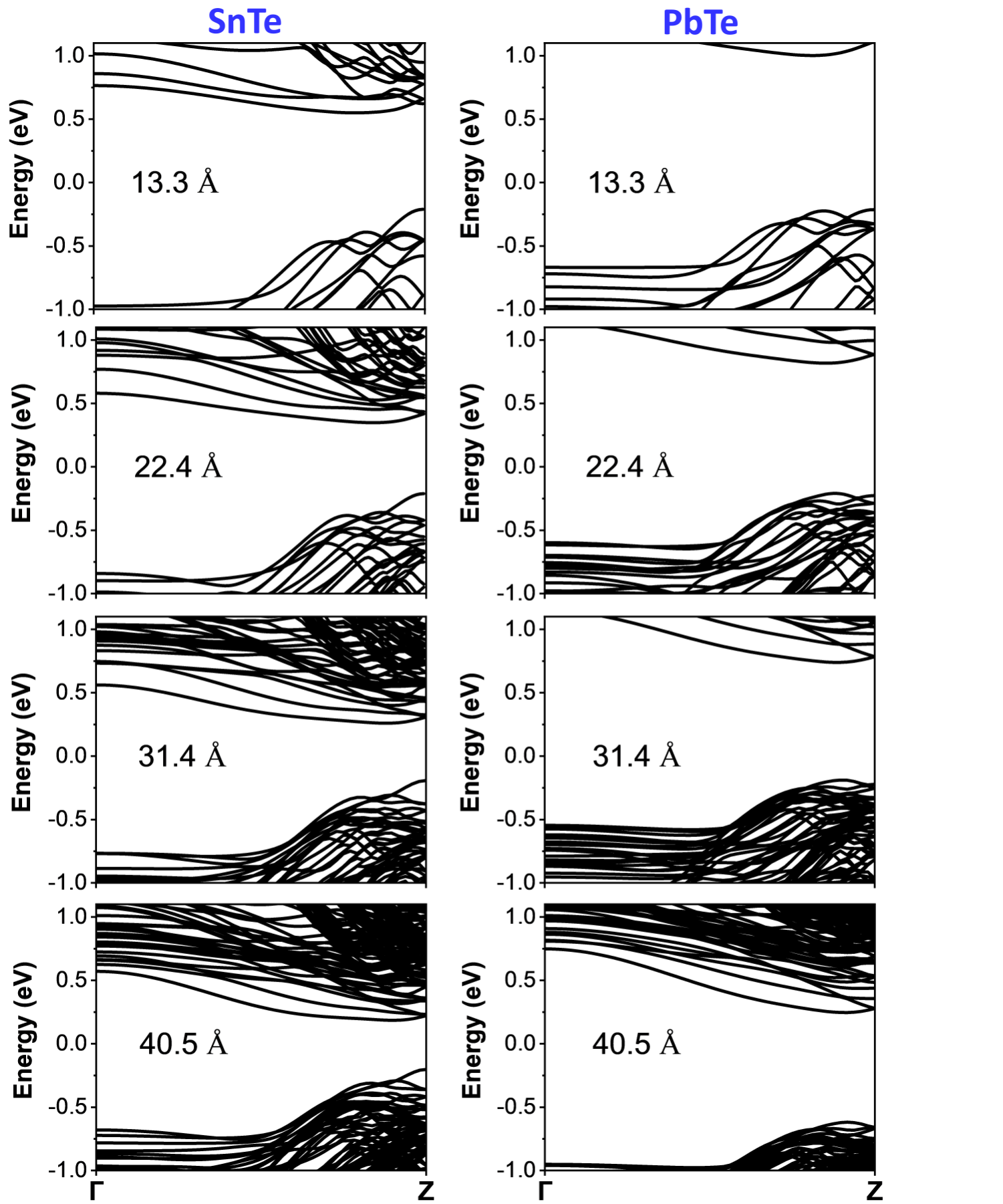}
\caption{Thickness-dependent electronic band structures of SnTe (left panel) and PbTe (right panel) cubic NWs grown along [001] without SOC. All the cubic nanowires reveal trivial insulating behavior. The Fermi level is set to zero. The conduction bands become closer and closer to the Fermi level as the thickness increases, which decreases the band gap. The thickness is shown in {\AA}, and the conversion to the size is reported in Table \ref{Table2}. Calculations were performed using the VASP code.}
\label{Band structures cubic}
\end{figure} 

Ultrathin nanowires show distortions from the NaCl phase. The phonon dispersion of the perfectly cubic ultrathin nanowires shows in Fig. \ref{quadruplar}a) negative frequencies and imaginary phonon modes. Once we include these distortions, all the phonon frequencies of the system become positive as we can see in Fig. \ref{quadruplar}b).
The finite-size effects produce an electric quadrupole with the Sn atoms attracting and the Te repelling each other as in Fig. \ref{quadruplar}c).
We have the clear formation of an electric quadruple with positive and negative charges that are slightly different due to the different atomic numbers, this produces non-zero forces on the atoms and displacements along the diagonals. This is very evident in the simple case of the 2x2 nanowires as shown in Fig. \ref{quadruplar}c). 
The quadrupolar distortions slightly affect the electronic and structural properties at low thicknesses.
When we increase the thickness, these quadrupolar distortions are still present on the surfaces of the nanowires while being reduced in the inner parts of the nanowire. The quadrupolar distortions do not break the inversion, mirror, and non-symmorphic symmetries.

In Figs. \ref{Bond length}a)-b), we show the bond lengths as a function of the thickness for both the SnTe and PbTe cubic nanowires oriented along the [001] direction. We can see that the bond lengths become closer to the theoretical values without relaxation when we go to larger thicknesses.
Therefore, these quadrupolar distortions are less and less relevant for thicker nanowires and they will not be considered in the study of the topology.
Also experimentally, it was reported that polar distortions in thin films are more relevant at low thicknesses\cite{Okamura23}.

\begin{table}[th]
\centering
\setlength{\tabcolsep}{8 pt}
\renewcommand{\arraystretch}{1.4}
\scalebox{1.2}{
\begin{tabular}{|c|c|c|c|}
\hline
Size & $t$  & Gap SnTe & Gap PbTe\\
\hline
   4x4 &  1.33   & 0.762  & 1.215 \\
   6x6 &  2.24   & 0.560  & 1.028 \\
   8x8 &  3.14   & 0.453  & 0.929 \\
 10x10 &  4.05   & 0.388  & 0.865  \\
\hline
\end{tabular}}
\caption{We report the size of the NW, the thicknesses $t$ (nm) equal for both SnTe and PbTe NWs, and the total band gaps (eV) from SOC calculations for SnTe and PbTe cubic nanowires after structural relaxation using the VASP code.}
\label{Table2}
\end{table}

Moving to the electronic properties, as all the structures are one-dimensional, we do not need to provide the density of states (DOS) since the band structure gives information about the entire Brillouin zone. We first present the electronic band structures without SOC of the SnTe and PbTe cubic NWs oriented along [001] direction in Fig. \ref{Band structures cubic}. All these NWs with size 10$\times$10 or smaller exhibit the trivial nature of insulators thereby revealing explicit band gaps. Going to the SOC calculations, Table \ref{Table2} displays the quantitative values of band gaps for SnTe and PbTe cubic NWs, with the band gap decreasing as a function of the thicknesses. As for the bulk systems, we show that also for nanowires the trivial band gap of PbTe is larger than the gap of SnTe. The alloying of Pb$_{1-x}$Sn$_x$Te would likely produce an intermediate gap between PbTe and SnTe. Moreover, the nature of the band gap is different for the two compounds i.e. SnTe cubic NWs show indirect band gaps, whereas direct band gaps appear for PbTe cubic NWs.

\begin{figure}[t]
\centering
\includegraphics[scale=0.23]{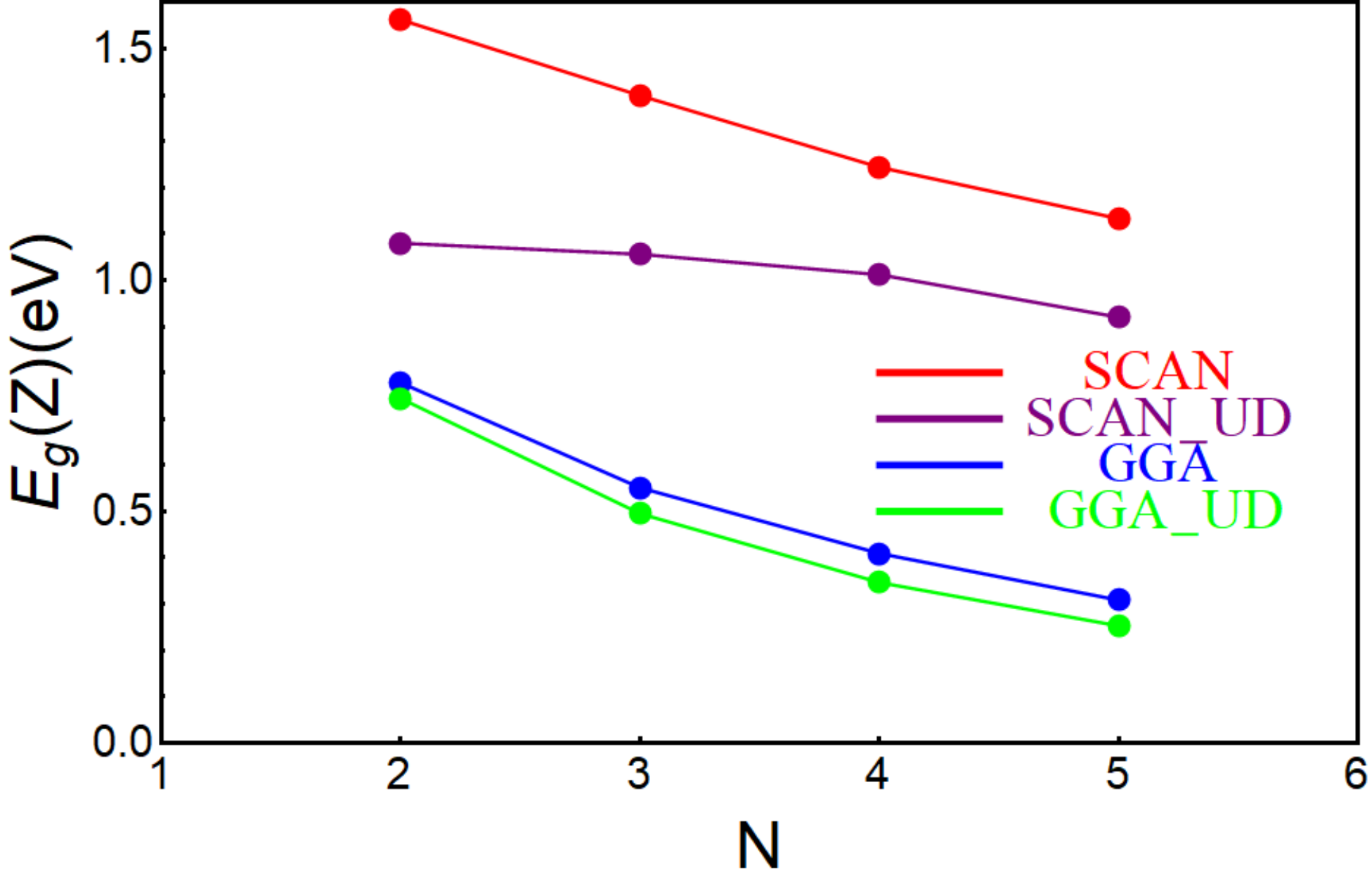}
\caption{Band gap of SnTe at Z point as a function of NWs thickness for GGA (blue) and SCAN (red). Undistorted (UD) nanowires band gaps are reported for GGA and SCAN in green and purple, respectively. The solid lines are a guide for the eyes. The in-plane number of atoms is 2N$\times$2N. Calculations were performed using the VASP code.}
\label{cubicgap_UD}
\end{figure}

\begin{figure}[t]
	\centering
\includegraphics[scale=0.34]{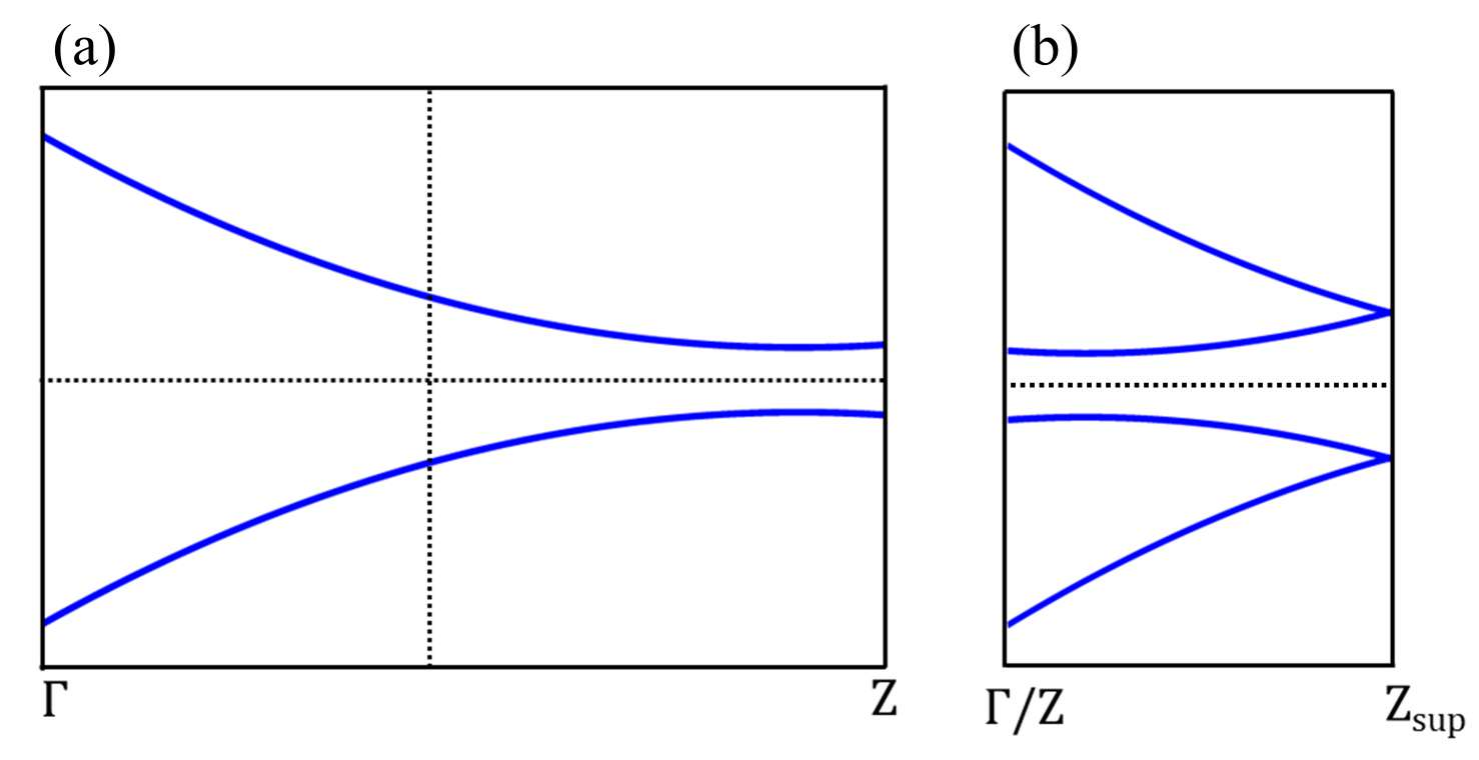}
\caption{a) Schematic band structure of one-dimensional TCI with the surface Dirac point close to the Z point with Z=$\frac{\pi}{c}$. The vertical dashed line represents half of the Brillouin zone. b) Schematic band structure of one-dimensional TCI with the folding of the $k$-path (due to the doubling of the real space unit cell). The Z point is folded in $\Gamma$, the surface Dirac point moves close to $\Gamma$. The new Brillouin zone ranges from $\Gamma$ to the Z of the supercell (Z$_{sup}$=$\frac{\pi}{2c}$) that is half of the initial Z. The horizontal dashed line is the Fermi level set to zero.}
\label{folding}
\end{figure}

\begin{figure*}[t]
\centering
\includegraphics[width=2\columnwidth,angle=0]{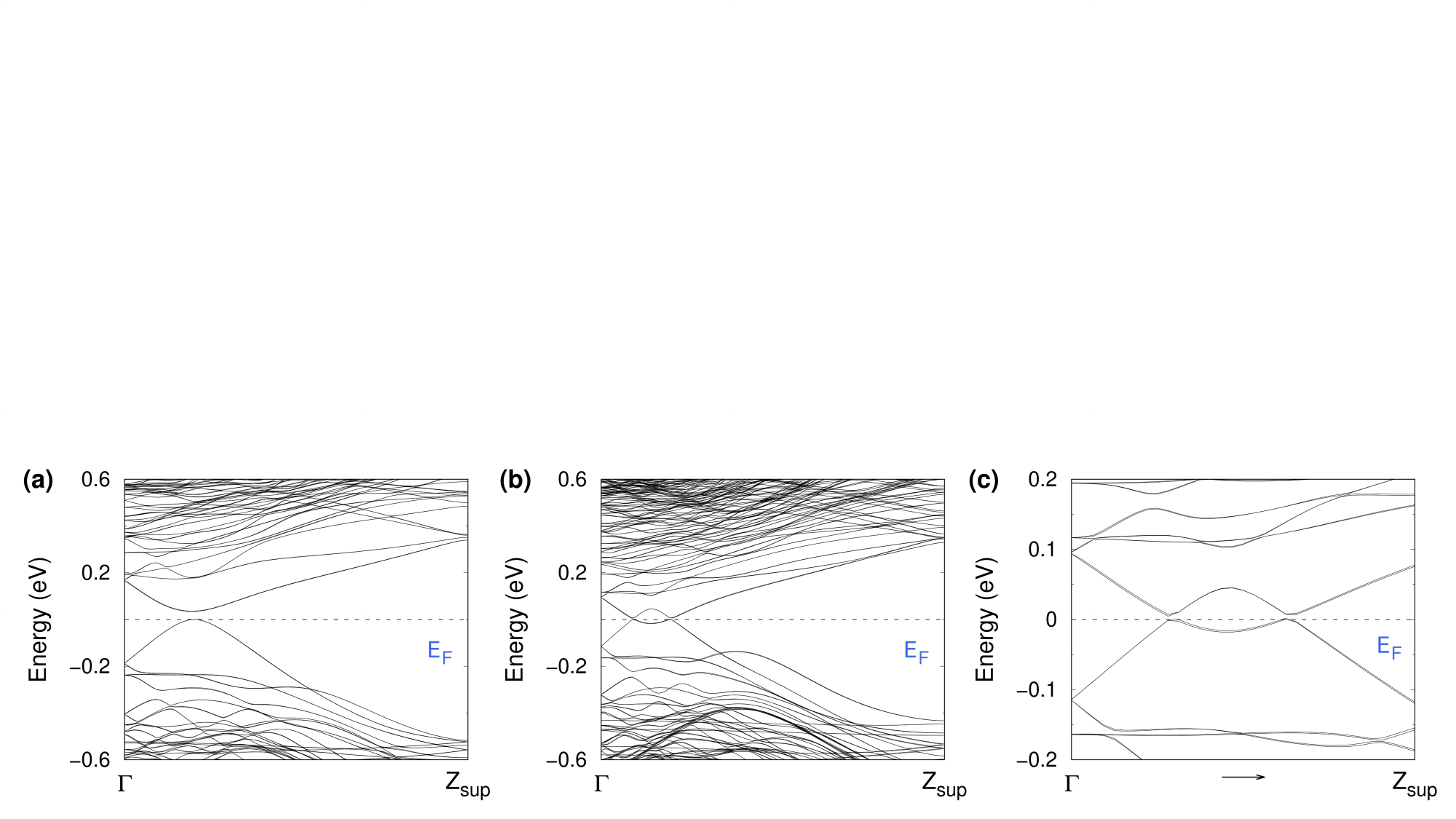}
\caption{Band structure of the SnTe nanowires with different sizes. a) Band gap insulating phase for the 10$\times$10$\times$4 NW, b) spin-orbit insulator for the 14$\times$14$\times$4 NW and c) magnification of the 14$\times$14$\times$4 NW. The calculations have been performed within ONETEP, the Brillouin zone ranges from $\Gamma$ to Z$_{sup}$. The Fermi level ($E_{\textnormal{F}}$) is set to zero.}
\label{soc_insulator}
\end{figure*}

\begin{figure}[t]
	\centering
\includegraphics[scale=0.315]{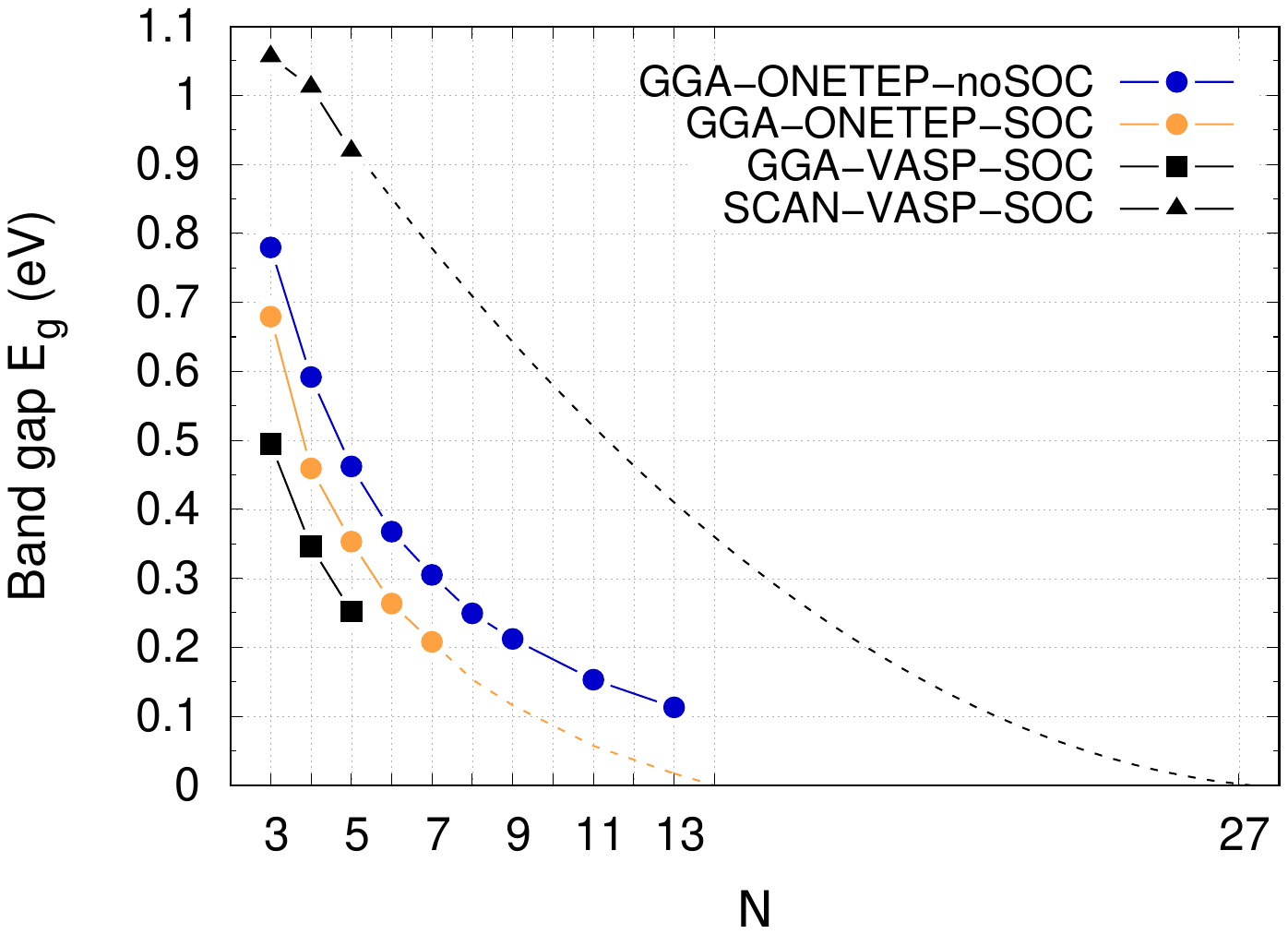}
\caption{Band gap of SnTe at $\Gamma$ point as a function of NWs thickness without and with SOC. The in-plane number of atoms is 2N$\times$2N. Calculations were performed using the ONETEP code with four atoms in the unit cell along the c-axis. The gap at the $\Gamma$ point with four atoms along the c-axis is equivalent to the gap at the Z-point reported in Fig. \ref{cubicgap_UD}. Blue points represent the gap without SOC, while yellow points represent the gap with SOC. The square points represent the band gap calculated with VASP without distortions. The solid lines are a guide for the eyes. We extrapolate the band gap going to zero from a parabolic fitting of the ONETEP data with SOC. The extrapolation of the SCAN data was performed with a linear fitting for gaps above 0.4 eV and a parabolic trend for the gap below 0.4 eV as from the band gap without SOC.}
\label{cubicgap_UD_ONETEP}
\end{figure}

\section{Phase diagram of the cubic S\lowercase{n}T\lowercase{e} NW\lowercase{S} as a function of the thickness}

The topological crystalline phase and the associated mirror Chern number are protected by mirror symmetry with respect to the (110) plane. The bulk SnTe has 6 surface Dirac points in the Brillouin zone, and the (001) surface of 2D topological SnTe hosts 4 surface Dirac points at non-time-reversal-invariant momenta\cite{PhysRevB.88.235126}. When the 2D Brillouin zone is projected on the 1D Brillouin zone, two Dirac points are projected in $\Gamma$ and become gapped due to their hybridization. The other 2 surface Dirac points of the (001) topological NWs rise close to +Z and -Z. Due to the large number of electronic bands, we cannot perform the wannierization and calculate the Chern number as it was done for the quaternary alloy bulk\cite{Cuono22}. However, we can study the band gap closure to search for the surface Dirac cones in the band structure of the NWs.

Here, we investigate the band gap of the SnTe cubic NWs as a function of the thickness using GGA and SCAN metaGGA exchange functional correlations and with the inclusion of the SOC within VASP (see Fig. \ref{cubicgap_UD}). We study the band gap at the high-symmetry points that are the relevant points for the topological properties. As the thickness of the NWs increases, the band gap decreases for both the functionals considered.
In the metaGGA approximation, the gap is larger and the system is more trivial.
For all the values of the thicknesses investigated within VASP, the system is trivial in both approximations.
To check the effects of the structural relaxation on the band gap, we also calculated the band gap for the undistorted system, namely without performing structural relaxation.
Although the gap without structural distortions is lower in energy than the gap of the distorted structure, in all cases analyzed the system is trivial.

In order to increase the thickness of the SnTe nanowires, we need to study larger supercells and we approach this using the linear scaling ONETEP code\cite{10.1063/1.1839852}. Due to technical reasons, within the ONETEP code, we are forced to double the structure along the z-axis. This doubling in the real space produced a halving of the first Brillouin zone in the $k$-space. The high-symmetry point Z [see Fig. \ref{folding}a)] is folded at the $\Gamma$ point, therefore, the 
band closing present close to Z moves close to $\Gamma$. 
The Brillouin zone of the supercell extends up to Z of the supercell (Z$_{sup}$). This scheme is shown in Fig. \ref{folding}b). Therefore, the band gap at the point $\Gamma$ within the supercell calculations (ONETEP) must be compared with the band gap at Z of the unit cell (VASP).

We plot in Fig. \ref{soc_insulator} the whole band structure for the two representative cases, namely the 10$\times$10$\times$4, the 14$\times$14$\times$4 NWs and its magnification. 
Fig. \ref{soc_insulator}a) shows the band insulator phase that appears in the ultrathin limit for the 10$\times$10$\times$4 NW or thinner. Increasing the size, the cubic SnTe NWs show some trivial surface states that fill the gap as shown in \ref{soc_insulator}b).  The spin-orbit opens a few meV gap between the valence and conduction bands creating an insulator as shown in Fig. \ref{soc_insulator}c), since the gap was not created at a high-symmetry point this is not a topological insulating phase but a trivial spin-orbit insulating phase. Considering the folding, the position of the trivial surface state is around the $k$-point 0.90$\frac{\pi}{c}$ in the unfolded Brillouin zone.  
When the band gap at the high-symmetry point reaches around 0.25 eV, we have the first transition from band gap insulator to spin-orbit insulator. 
These trivial surface states appear on a wider range of thicknesses with respect to the topological surface states, therefore, they are quite robust even if not topologically protected. 
These trivial surface states lie at the same energy and $k$-point where the topological surface states should form the Dirac point. Since these trivial surface states would persist in the topological phase, we would have a co-presence of trivial and topological surface states at around 0.90$\frac{\pi}{c}$. 
The presence of these trivial bands would increase the number of carriers in SnTe and in general in 1D Pb$_{1-x}$Sn$_x$Te systems. These electronic and topological properties are strongly dependent on the nanowire's thickness as experimentally confirmed\cite{C4NR05870D}.

The \textit{ab initio} calculations will be closer to the experimental materials with respect to the model Hamiltonian where the same SOC term and the same hopping parameters are used for Sn and Te atoms. Within \textit{ab initio} calculations, we find these additional trivial surface states that are present also in the model Hamiltonian calculations\cite{Nguyen22,10.21468/SciPostPhysCore.6.1.011}, however, the spin-orbit gap is much smaller with DFT.
Therefore, the presence of trivial surface bands at the Fermi level beyond the surface Dirac bands is not an artifact of the application of the 3D model Hamiltonian to the 1D system but it is a real property of the IV-VI NWs. These trivial surface states hybridize with the surface Dirac points since they are at the same energy and the same $k$-point; this hybridization produces massive Dirac points\cite{Rechcinski21}. To have a closed Dirac point, we need to go to very high thicknesses \cite{Schindler18} to avoid hybridization between the surfaces.
Some elements of high-order topology were reported in model Hamiltonian papers even without breaking the mirror symmetry, indeed, the hinge states in cubic NWs are present for large thicknesses at the point Z\cite{Nguyen22} or very close to Z\cite{10.21468/SciPostPhysCore.6.1.011}.

In Fig. \ref{cubicgap_UD_ONETEP}, we report the band gap of SnTe at $\Gamma$ point as a function of the film thickness without and with SOC by using the supercell with 4 atoms along the c-axis and the ONETEP code. We compare it with the band gap calculated with VASP within GGA and SCAN. In all cases, the gaps go to zero increasing the number of atoms and therefore the number of hoppings and the total bandwidth.
Examining the ONETEP calculations with and without SOC,
we observe that the band gap becomes smaller with SOC due to the band splitting in the valence band and conduction band. 
Increasing N, the reduction of the gap becomes lower and lower, this is due to the increasing hybridization at the $\Gamma$ point between the topmost valence band and the lowest conduction band. The quantitative difference between VASP and ONETEP is attributed to different implementations and approximations of the two DFT codes. The SCAN results are expected to be the ones closer to the experimental results, we extrapolate the SCAN results in order to obtain realistic values for the topological transition thicknesses. 

\begin{figure*}[t]
\centering
\includegraphics[width=2\columnwidth,angle=0]{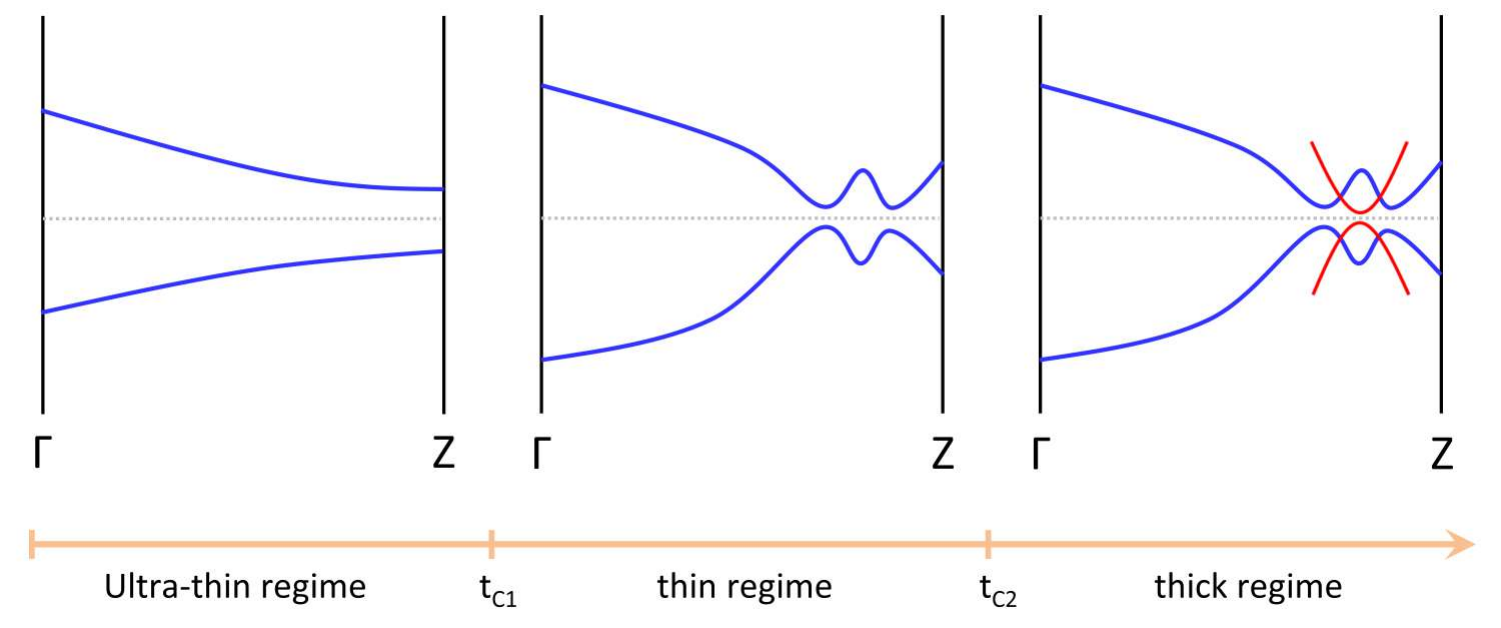}
\caption{Phase diagram as a function of the thickness $t$ for cubic Pb$_{1-x}$Sn$_x$Te NWs. The different phases are separated by the critical thickness values t$_{c1}$ and t$_{c2}$. The blue bands are trivial, while the surface massive Dirac fermions are plotted in red. We have a co-presence of trivial and topological surface states at around 0.90$\frac{\pi}{c}$. The horizontal dashed lines represent the Fermi level. Extrapolating the data from SCAN calculations, realistic critical values for pure cubic SnTe NWs are t$_{c1}$=10 nm and t$_{c2}$=17 nm. We expect the same phase diagram for Pb$_{1-x}$Sn$_x$Te NWs but with larger values of t$_{c1}$ and t$_{c2}$ depending on the doping concentration.}
\label{phase_diagram}
\end{figure*} 

Within the ONETEP calculations, the first transition from band insulator to SOC-insulator happens at N=6 when the gap at $\Gamma$ goes below 0.3 eV, while the topological transition happens at N=14 when the band gap of the system with SOC goes to zero according with the extrapolation of Fig. \ref{cubicgap_UD_ONETEP}. 
The critical thickness of the nanowires is given by $t=(2N-1)d_{SnTe}$, where $d_{SnTe}$ is the distance between Sn and Te which is half of the lattice constant. 
From this formula, we calculate the critical values and we obtain $t_{c1}^{GGA}$=3.5 nm and $t_{c2}^{GGA}$= 8.5 nm for the GGA ONETEP calculations.

However, we know that a more realistic value for the topological transition would be within the metaGGCA SCAN. We extrapolate from the SCAN data the behavior up to 0.4 eV, below 0.4 we assume that the reduction of the gap is slowed down by the hybridization between conduction and valence bands as we can see in the ONETEP results without SOC.
Therefore, we assume that behavior for the gap. From our assumptions, we get the first transition at N=16 and the second transition at N=27.
Since SCAN is good also for surfaces\cite{10.1063/1.5134951}, we start from the SCAN results and extrapolate the data adding the behavior of the NOSOC that we were able to send to zero.
We obtain that the system reaches the gap of circa 0.25 eV for N= 16 which corresponds to t$_{c1}$=10 nm, while the critical value of the topological transition is N=27 which is equivalent to t$_{c2}$=17 nm.
The trivial surface states appear in a wider range than the topological surface states; SnTe NWS with thicknesses 17 nm or below were synthesized\cite{Delft_thesis,Sadowski2018}, therefore, we conclude that these NWs are not topological but they are in the SOC insulating phase.

\section{Discussion and Conclusions}    

In order to investigate Majorana fermions, the SnTe nanowires should host superconductivity beyond the topological properties demonstrated in the previous Section. The superconductivity can be realized using proximity effect\cite{Klett18}, by doping \cite{Mitobe21} or interfacing SnTe with PbTe \cite{Fogel01,Tang14}.

The possibility of ferroelectricity of the cubic phase is addressed in a recent experimental paper\cite{liu2020synthesis}, further studies are necessary on ferroelectricity and its interplay with the topology. In SnTe, the ferroelectricity produces the Rashba effect because the inversion symmetry is broken; in PbTe we have strong SOC, then in Pb$_{1-x}$Sn$_x$Te, due to the combination of lack of inversion symmetry and strong SOC, we can have ferroelectric Rashba topological system\cite{PhysRevB.108.045114,nano10040732}.

In conclusion, we have shown using the DFT that the thin cubic SnTe and PbTe nanowires are trivial insulators in the ultrathin limit. 
A quadrupolar distortion relevant on the surface of the NWs slightly increases the band gap. 
At the same thicknesses, the PbTe NWs have a larger trivial gap than SnTe NWs.
The cubic SnTe NWs undergo an electronic transition from the band gap insulator 
to a spin-orbit insulator at the thickness t$_{c1}$=10 nm. Increasing again the thickness, the cubic SnTe NWs become topological crystalline insulators for a thickness larger than t$_{c1}$=17 nm. In the topological phase, the SnTe NWs host two massive Dirac surface states at non time-reversal invariant momenta close to the Z point hybridizing with trivial surface states.
The SnTe cubic nanowires experimentally reported by different authors with thicknesses ranging between 10 and 200 nm\cite{safdar2013topological,Sadowski2018,liu2020synthesis,Delft_thesis} are mainly topological systems with the exception of the thin NWs with a thickness smaller than t$_{c2}$=17 nm.
The IV-VI NWs are topologically more trivial than the IV-VI bulk systems, therefore the Sn-concentration necessary to activate the topological transition in Pb$_{1-x}$Sn$_x$Te will be higher than the bulk values and dependent on the thickness of the NWs. Moving to Pb$_{1-x}$Sn$_x$Te NWs, the values of t$_{c1}$ and t$_{c2}$ could sensibly increase especially close to the level of doping that induces the topological transition in the bulk.
Since experimental SnTe cubic NWs are topological and could host superconductivity as well, we conclude that these systems are a suitable platform for the investigation of Majorana fermions.

\vspace{0.1cm}

\begin{acknowledgments}
G. H. and K. W. contributed equally to this work. 
The authors thank Tomasz Dietl for encouraging us to work on topological nanowires. We acknowledge  R. Buczko, S. Samadi, A. Kazakov, W. Brzezicki, C. M. Canali and S. Sattar for constructive discussions. K. W. acknowledges S. Winczewski and J. Dziedzic for useful suggestions about the use of the ONETEP code.
The work is supported by the Foundation for Polish Science through the International Research Agendas program co-financed by the European Union within the Smart Growth Operational Programme (Grant No. MAB/2017/1). 
We acknowledge the access to the computing facilities of the Interdisciplinary Center of Modeling at the University of Warsaw, Grant g91-1418, g91-1419 and g91-1426 for the availability of high-performance computing resources and support.
We acknowledge the CINECA award under the ISCRA initiative  IsC99 "SILENTS”, IsC105 "SILENTSG" and IsB26 "SHINY" grants for the availability of high-performance computing resources and support. We acknowledge the access to the computing facilities of the Poznan Supercomputing and Networking Center Grants No. 609.\\
\end{acknowledgments}

\bibliography{bibliography}

\begin{thebibliography}{77}%
\makeatletter
\providecommand \@ifxundefined [1]{%
 \@ifx{#1\undefined}
}%
\providecommand \@ifnum [1]{%
 \ifnum #1\expandafter \@firstoftwo
 \else \expandafter \@secondoftwo
 \fi
}%
\providecommand \@ifx [1]{%
 \ifx #1\expandafter \@firstoftwo
 \else \expandafter \@secondoftwo
 \fi
}%
\providecommand \natexlab [1]{#1}%
\providecommand \enquote  [1]{``#1''}%
\providecommand \bibnamefont  [1]{#1}%
\providecommand \bibfnamefont [1]{#1}%
\providecommand \citenamefont [1]{#1}%
\providecommand \href@noop [0]{\@secondoftwo}%
\providecommand \href [0]{\begingroup \@sanitize@url \@href}%
\providecommand \@href[1]{\@@startlink{#1}\@@href}%
\providecommand \@@href[1]{\endgroup#1\@@endlink}%
\providecommand \@sanitize@url [0]{\catcode `\\12\catcode `\$12\catcode
  `\&12\catcode `\#12\catcode `\^12\catcode `\_12\catcode `\%12\relax}%
\providecommand \@@startlink[1]{}%
\providecommand \@@endlink[0]{}%
\providecommand \url  [0]{\begingroup\@sanitize@url \@url }%
\providecommand \@url [1]{\endgroup\@href {#1}{\urlprefix }}%
\providecommand \urlprefix  [0]{URL }%
\providecommand \Eprint [0]{\href }%
\providecommand \doibase [0]{http://dx.doi.org/}%
\providecommand \selectlanguage [0]{\@gobble}%
\providecommand \bibinfo  [0]{\@secondoftwo}%
\providecommand \bibfield  [0]{\@secondoftwo}%
\providecommand \translation [1]{[#1]}%
\providecommand \BibitemOpen [0]{}%
\providecommand \bibitemStop [0]{}%
\providecommand \bibitemNoStop [0]{.\EOS\space}%
\providecommand \EOS [0]{\spacefactor3000\relax}%
\providecommand \BibitemShut  [1]{\csname bibitem#1\endcsname}%
\let\auto@bib@innerbib\@empty
\bibitem [{\citenamefont {Sau}\ \emph {et~al.}(2020)\citenamefont {Sau},
  \citenamefont {Simon}, \citenamefont {Vishveshwara},\ and\ \citenamefont
  {Williams}}]{Sau2020}%
  \BibitemOpen
  \bibfield  {author} {\bibinfo {author} {\bibfnamefont {J.}~\bibnamefont
  {Sau}}, \bibinfo {author} {\bibfnamefont {S.}~\bibnamefont {Simon}}, \bibinfo
  {author} {\bibfnamefont {S.}~\bibnamefont {Vishveshwara}}, \ and\ \bibinfo
  {author} {\bibfnamefont {J.~R.}\ \bibnamefont {Williams}},\ }\href {\doibase
  10.1038/s42254-020-00251-9} {\bibfield  {journal} {\bibinfo  {journal}
  {Nature Reviews Physics}\ }\textbf {\bibinfo {volume} {2}},\ \bibinfo {pages}
  {667} (\bibinfo {year} {2020})}\BibitemShut {NoStop}%
\bibitem [{\citenamefont {Beenakker}(2013)}]{Beenakker2013}%
  \BibitemOpen
  \bibfield  {author} {\bibinfo {author} {\bibfnamefont {C.}~\bibnamefont
  {Beenakker}},\ }\href {\doibase 10.1146/annurev-conmatphys-030212-184337}
  {\bibfield  {journal} {\bibinfo  {journal} {Annual Review of Condensed Matter
  Physics}\ }\textbf {\bibinfo {volume} {4}},\ \bibinfo {pages} {113} (\bibinfo
  {year} {2013})},\ \Eprint
  {http://arxiv.org/abs/https://doi.org/10.1146/annurev-conmatphys-030212-184337}
  {https://doi.org/10.1146/annurev-conmatphys-030212-184337} \BibitemShut
  {NoStop}%
\bibitem [{\citenamefont {Frolov}(2021)}]{Frolov21}%
  \BibitemOpen
  \bibfield  {author} {\bibinfo {author} {\bibfnamefont {S.}~\bibnamefont
  {Frolov}},\ }\href {\doibase 10.1038/d41586-021-00954-8} {\bibfield
  {journal} {\bibinfo  {journal} {Nature}\ }\textbf {\bibinfo {volume} {592}},\
  \bibinfo {pages} {350} (\bibinfo {year} {2021})}\BibitemShut {NoStop}%
\bibitem [{\citenamefont {Stanescu}\ and\ \citenamefont
  {Tewari}(2013)}]{Stanescu2013}%
  \BibitemOpen
  \bibfield  {author} {\bibinfo {author} {\bibfnamefont {T.~D.}\ \bibnamefont
  {Stanescu}}\ and\ \bibinfo {author} {\bibfnamefont {S.}~\bibnamefont
  {Tewari}},\ }\href {\doibase 10.1088/0953-8984/25/23/233201} {\bibfield
  {journal} {\bibinfo  {journal} {Journal of Physics: Condensed Matter}\
  }\textbf {\bibinfo {volume} {25}},\ \bibinfo {pages} {233201} (\bibinfo
  {year} {2013})}\BibitemShut {NoStop}%
\bibitem [{\citenamefont {Wang}\ \emph {et~al.}(2011)\citenamefont {Wang},
  \citenamefont {Pei}, \citenamefont {LaLonde},\ and\ \citenamefont
  {Snyder}}]{wang2011heavily}%
  \BibitemOpen
  \bibfield  {author} {\bibinfo {author} {\bibfnamefont {H.}~\bibnamefont
  {Wang}}, \bibinfo {author} {\bibfnamefont {Y.}~\bibnamefont {Pei}}, \bibinfo
  {author} {\bibfnamefont {A.~D.}\ \bibnamefont {LaLonde}}, \ and\ \bibinfo
  {author} {\bibfnamefont {G.~J.}\ \bibnamefont {Snyder}},\ }\href@noop {}
  {\bibfield  {journal} {\bibinfo  {journal} {Advanced Materials}\ }\textbf
  {\bibinfo {volume} {23}},\ \bibinfo {pages} {1366} (\bibinfo {year}
  {2011})}\BibitemShut {NoStop}%
\bibitem [{\citenamefont {Wood}(1988)}]{wood1988materials}%
  \BibitemOpen
  \bibfield  {author} {\bibinfo {author} {\bibfnamefont {C.}~\bibnamefont
  {Wood}},\ }\href@noop {} {\bibfield  {journal} {\bibinfo  {journal} {Reports
  on progress in physics}\ }\textbf {\bibinfo {volume} {51}},\ \bibinfo {pages}
  {459} (\bibinfo {year} {1988})}\BibitemShut {NoStop}%
\bibitem [{\citenamefont {Lebedev}\ and\ \citenamefont
  {Sluchinskaya}(1994)}]{lebedev1994ferroelectric}%
  \BibitemOpen
  \bibfield  {author} {\bibinfo {author} {\bibfnamefont {A.~I.}\ \bibnamefont
  {Lebedev}}\ and\ \bibinfo {author} {\bibfnamefont {I.~A.}\ \bibnamefont
  {Sluchinskaya}},\ }\href@noop {} {\bibfield  {journal} {\bibinfo  {journal}
  {Ferroelectrics}\ }\textbf {\bibinfo {volume} {157}},\ \bibinfo {pages} {275}
  (\bibinfo {year} {1994})}\BibitemShut {NoStop}%
\bibitem [{\citenamefont {Liu}\ \emph {et~al.}(2021)\citenamefont {Liu},
  \citenamefont {Han}, \citenamefont {Wei}, \citenamefont {Hynek},
  \citenamefont {Hart}, \citenamefont {Han}, \citenamefont {Trimble},
  \citenamefont {Williams}, \citenamefont {Zhu},\ and\ \citenamefont
  {Cha}}]{liu2020synthesis}%
  \BibitemOpen
  \bibfield  {author} {\bibinfo {author} {\bibfnamefont {P.}~\bibnamefont
  {Liu}}, \bibinfo {author} {\bibfnamefont {H.~J.}\ \bibnamefont {Han}},
  \bibinfo {author} {\bibfnamefont {J.}~\bibnamefont {Wei}}, \bibinfo {author}
  {\bibfnamefont {D.~J.}\ \bibnamefont {Hynek}}, \bibinfo {author}
  {\bibfnamefont {J.~L.}\ \bibnamefont {Hart}}, \bibinfo {author}
  {\bibfnamefont {M.~G.}\ \bibnamefont {Han}}, \bibinfo {author} {\bibfnamefont
  {C.~J.}\ \bibnamefont {Trimble}}, \bibinfo {author} {\bibfnamefont {J.~R.}\
  \bibnamefont {Williams}}, \bibinfo {author} {\bibfnamefont {Y.}~\bibnamefont
  {Zhu}}, \ and\ \bibinfo {author} {\bibfnamefont {J.~J.}\ \bibnamefont
  {Cha}},\ }\href {\doibase 10.1021/acsaelm.0c00740} {\bibfield  {journal}
  {\bibinfo  {journal} {ACS Applied Electronic Materials}\ }\textbf {\bibinfo
  {volume} {3}},\ \bibinfo {pages} {184} (\bibinfo {year} {2021})}\BibitemShut
  {NoStop}%
\bibitem [{\citenamefont {Matsushita}\ \emph {et~al.}(2006)\citenamefont
  {Matsushita}, \citenamefont {Wianecki}, \citenamefont {Sommer}, \citenamefont
  {Geballe},\ and\ \citenamefont {Fisher}}]{matsushita2006type}%
  \BibitemOpen
  \bibfield  {author} {\bibinfo {author} {\bibfnamefont {Y.}~\bibnamefont
  {Matsushita}}, \bibinfo {author} {\bibfnamefont {P.}~\bibnamefont
  {Wianecki}}, \bibinfo {author} {\bibfnamefont {A.~T.}\ \bibnamefont
  {Sommer}}, \bibinfo {author} {\bibfnamefont {T.}~\bibnamefont {Geballe}}, \
  and\ \bibinfo {author} {\bibfnamefont {I.}~\bibnamefont {Fisher}},\
  }\href@noop {} {\bibfield  {journal} {\bibinfo  {journal} {Physical Review
  B}\ }\textbf {\bibinfo {volume} {74}},\ \bibinfo {pages} {134512} (\bibinfo
  {year} {2006})}\BibitemShut {NoStop}%
\bibitem [{\citenamefont {Mazur}\ \emph {et~al.}(2019)\citenamefont {Mazur},
  \citenamefont {Dybko}, \citenamefont {Szczerbakow}, \citenamefont {Domagala},
  \citenamefont {Kazakov}, \citenamefont {Zgirski}, \citenamefont {Lusakowska},
  \citenamefont {Kret}, \citenamefont {Korczak}, \citenamefont {Story},
  \citenamefont {Sawicki},\ and\ \citenamefont
  {Dietl}}]{mazur2019experimental}%
  \BibitemOpen
  \bibfield  {author} {\bibinfo {author} {\bibfnamefont {G.~P.}\ \bibnamefont
  {Mazur}}, \bibinfo {author} {\bibfnamefont {K.}~\bibnamefont {Dybko}},
  \bibinfo {author} {\bibfnamefont {A.}~\bibnamefont {Szczerbakow}}, \bibinfo
  {author} {\bibfnamefont {J.~Z.}\ \bibnamefont {Domagala}}, \bibinfo {author}
  {\bibfnamefont {A.}~\bibnamefont {Kazakov}}, \bibinfo {author} {\bibfnamefont
  {M.}~\bibnamefont {Zgirski}}, \bibinfo {author} {\bibfnamefont
  {E.}~\bibnamefont {Lusakowska}}, \bibinfo {author} {\bibfnamefont
  {S.}~\bibnamefont {Kret}}, \bibinfo {author} {\bibfnamefont {J.}~\bibnamefont
  {Korczak}}, \bibinfo {author} {\bibfnamefont {T.}~\bibnamefont {Story}},
  \bibinfo {author} {\bibfnamefont {M.}~\bibnamefont {Sawicki}}, \ and\
  \bibinfo {author} {\bibfnamefont {T.}~\bibnamefont {Dietl}},\ }\href
  {\doibase 10.1103/PhysRevB.100.041408} {\bibfield  {journal} {\bibinfo
  {journal} {Phys. Rev. B}\ }\textbf {\bibinfo {volume} {100}},\ \bibinfo
  {pages} {041408} (\bibinfo {year} {2019})}\BibitemShut {NoStop}%
\bibitem [{\citenamefont {Fu}(2011)}]{fu2011topological}%
  \BibitemOpen
  \bibfield  {author} {\bibinfo {author} {\bibfnamefont {L.}~\bibnamefont
  {Fu}},\ }\href@noop {} {\bibfield  {journal} {\bibinfo  {journal} {Physical
  Review Letters}\ }\textbf {\bibinfo {volume} {106}},\ \bibinfo {pages}
  {106802} (\bibinfo {year} {2011})}\BibitemShut {NoStop}%
\bibitem [{\citenamefont {Hsieh}\ \emph {et~al.}(2012)\citenamefont {Hsieh},
  \citenamefont {Lin}, \citenamefont {Liu}, \citenamefont {Duan}, \citenamefont
  {Bansil},\ and\ \citenamefont {Fu}}]{hsieh2012topological}%
  \BibitemOpen
  \bibfield  {author} {\bibinfo {author} {\bibfnamefont {T.~H.}\ \bibnamefont
  {Hsieh}}, \bibinfo {author} {\bibfnamefont {H.}~\bibnamefont {Lin}}, \bibinfo
  {author} {\bibfnamefont {J.}~\bibnamefont {Liu}}, \bibinfo {author}
  {\bibfnamefont {W.}~\bibnamefont {Duan}}, \bibinfo {author} {\bibfnamefont
  {A.}~\bibnamefont {Bansil}}, \ and\ \bibinfo {author} {\bibfnamefont
  {L.}~\bibnamefont {Fu}},\ }\href@noop {} {\bibfield  {journal} {\bibinfo
  {journal} {Nature communications}\ }\textbf {\bibinfo {volume} {3}},\
  \bibinfo {pages} {1} (\bibinfo {year} {2012})}\BibitemShut {NoStop}%
\bibitem [{\citenamefont {Lau}\ and\ \citenamefont {Ortix}(2019)}]{Lau19}%
  \BibitemOpen
  \bibfield  {author} {\bibinfo {author} {\bibfnamefont {A.}~\bibnamefont
  {Lau}}\ and\ \bibinfo {author} {\bibfnamefont {C.}~\bibnamefont {Ortix}},\
  }\href {\doibase 10.1103/PhysRevLett.122.186801} {\bibfield  {journal}
  {\bibinfo  {journal} {Phys. Rev. Lett.}\ }\textbf {\bibinfo {volume} {122}},\
  \bibinfo {pages} {186801} (\bibinfo {year} {2019})}\BibitemShut {NoStop}%
\bibitem [{\citenamefont {Tanaka}\ \emph {et~al.}(2012)\citenamefont {Tanaka},
  \citenamefont {Ren}, \citenamefont {Sato}, \citenamefont {Nakayama},
  \citenamefont {Souma}, \citenamefont {Takahashi}, \citenamefont {Segawa},\
  and\ \citenamefont {Ando}}]{Tanaka07}%
  \BibitemOpen
  \bibfield  {author} {\bibinfo {author} {\bibfnamefont {Y.}~\bibnamefont
  {Tanaka}}, \bibinfo {author} {\bibfnamefont {Z.}~\bibnamefont {Ren}},
  \bibinfo {author} {\bibfnamefont {T.}~\bibnamefont {Sato}}, \bibinfo {author}
  {\bibfnamefont {K.}~\bibnamefont {Nakayama}}, \bibinfo {author}
  {\bibfnamefont {T.}~\bibnamefont {Souma}, \bibfnamefont {S.~a nd~Takahashi}},
  \bibinfo {author} {\bibfnamefont {T.}~\bibnamefont {Takahashi}}, \bibinfo
  {author} {\bibfnamefont {K.}~\bibnamefont {Segawa}}, \ and\ \bibinfo {author}
  {\bibfnamefont {Y.}~\bibnamefont {Ando}},\ }\href {\doibase
  10.1038/nphys2442} {\bibfield  {journal} {\bibinfo  {journal} {Nature
  Physics}\ }\textbf {\bibinfo {volume} {8}},\ \bibinfo {pages} {800} (\bibinfo
  {year} {2012})}\BibitemShut {NoStop}%
\bibitem [{\citenamefont {Okada}\ \emph {et~al.}(2013)\citenamefont {Okada},
  \citenamefont {Serbyn}, \citenamefont {Lin}, \citenamefont {Walkup},
  \citenamefont {Zhou}, \citenamefont {Dhital}, \citenamefont {Neupane},
  \citenamefont {Xu}, \citenamefont {Wang}, \citenamefont {Sankar},
  \citenamefont {Chou}, \citenamefont {Bansil}, \citenamefont {Hasan},
  \citenamefont {Wilson}, \citenamefont {Fu},\ and\ \citenamefont
  {Madhavan}}]{Okada13}%
  \BibitemOpen
  \bibfield  {author} {\bibinfo {author} {\bibfnamefont {Y.}~\bibnamefont
  {Okada}}, \bibinfo {author} {\bibfnamefont {M.}~\bibnamefont {Serbyn}},
  \bibinfo {author} {\bibfnamefont {H.}~\bibnamefont {Lin}}, \bibinfo {author}
  {\bibfnamefont {D.}~\bibnamefont {Walkup}}, \bibinfo {author} {\bibfnamefont
  {W.}~\bibnamefont {Zhou}}, \bibinfo {author} {\bibfnamefont {C.}~\bibnamefont
  {Dhital}}, \bibinfo {author} {\bibfnamefont {M.}~\bibnamefont {Neupane}},
  \bibinfo {author} {\bibfnamefont {S.}~\bibnamefont {Xu}}, \bibinfo {author}
  {\bibfnamefont {Y.~J.}\ \bibnamefont {Wang}}, \bibinfo {author}
  {\bibfnamefont {R.}~\bibnamefont {Sankar}}, \bibinfo {author} {\bibfnamefont
  {F.}~\bibnamefont {Chou}}, \bibinfo {author} {\bibfnamefont {A.}~\bibnamefont
  {Bansil}}, \bibinfo {author} {\bibfnamefont {M.~Z.}\ \bibnamefont {Hasan}},
  \bibinfo {author} {\bibfnamefont {S.~D.}\ \bibnamefont {Wilson}}, \bibinfo
  {author} {\bibfnamefont {L.}~\bibnamefont {Fu}}, \ and\ \bibinfo {author}
  {\bibfnamefont {V.}~\bibnamefont {Madhavan}},\ }\href {\doibase
  10.1126/science.1239451} {\bibfield  {journal} {\bibinfo  {journal}
  {Science}\ }\textbf {\bibinfo {volume} {341}},\ \bibinfo {pages} {1496}
  (\bibinfo {year} {2013})}\BibitemShut {NoStop}%
\bibitem [{\citenamefont {Xu}\ \emph {et~al.}(2012)\citenamefont {Xu},
  \citenamefont {Liu}, \citenamefont {Alidoust}, \citenamefont {Neupane},
  \citenamefont {Qian}, \citenamefont {Belopolski}, \citenamefont {Denlinger},
  \citenamefont {Wang}, \citenamefont {Lin}, \citenamefont {Wray} \emph
  {et~al.}}]{xu2012observation}%
  \BibitemOpen
  \bibfield  {author} {\bibinfo {author} {\bibfnamefont {S.-Y.}\ \bibnamefont
  {Xu}}, \bibinfo {author} {\bibfnamefont {C.}~\bibnamefont {Liu}}, \bibinfo
  {author} {\bibfnamefont {N.}~\bibnamefont {Alidoust}}, \bibinfo {author}
  {\bibfnamefont {M.}~\bibnamefont {Neupane}}, \bibinfo {author} {\bibfnamefont
  {D.}~\bibnamefont {Qian}}, \bibinfo {author} {\bibfnamefont {I.}~\bibnamefont
  {Belopolski}}, \bibinfo {author} {\bibfnamefont {J.}~\bibnamefont
  {Denlinger}}, \bibinfo {author} {\bibfnamefont {Y.}~\bibnamefont {Wang}},
  \bibinfo {author} {\bibfnamefont {H.}~\bibnamefont {Lin}}, \bibinfo {author}
  {\bibfnamefont {L.~a.}\ \bibnamefont {Wray}},  \emph {et~al.},\ }\href@noop
  {} {\bibfield  {journal} {\bibinfo  {journal} {Nature communications}\
  }\textbf {\bibinfo {volume} {3}},\ \bibinfo {pages} {1} (\bibinfo {year}
  {2012})}\BibitemShut {NoStop}%
\bibitem [{\citenamefont {Dziawa}\ \emph {et~al.}(2012)\citenamefont {Dziawa},
  \citenamefont {Kowalski}, \citenamefont {Dybko}, \citenamefont {Buczko},
  \citenamefont {Szczerbakow}, \citenamefont {Szot}, \citenamefont
  {{\L}usakowska}, \citenamefont {Balasubramanian}, \citenamefont {Wojek},
  \citenamefont {Berntsen} \emph {et~al.}}]{dziawa2012topological}%
  \BibitemOpen
  \bibfield  {author} {\bibinfo {author} {\bibfnamefont {P.}~\bibnamefont
  {Dziawa}}, \bibinfo {author} {\bibfnamefont {B.}~\bibnamefont {Kowalski}},
  \bibinfo {author} {\bibfnamefont {K.}~\bibnamefont {Dybko}}, \bibinfo
  {author} {\bibfnamefont {R.}~\bibnamefont {Buczko}}, \bibinfo {author}
  {\bibfnamefont {A.}~\bibnamefont {Szczerbakow}}, \bibinfo {author}
  {\bibfnamefont {M.}~\bibnamefont {Szot}}, \bibinfo {author} {\bibfnamefont
  {E.}~\bibnamefont {{\L}usakowska}}, \bibinfo {author} {\bibfnamefont
  {T.}~\bibnamefont {Balasubramanian}}, \bibinfo {author} {\bibfnamefont
  {B.~M.}\ \bibnamefont {Wojek}}, \bibinfo {author} {\bibfnamefont
  {M.}~\bibnamefont {Berntsen}},  \emph {et~al.},\ }\href@noop {} {\bibfield
  {journal} {\bibinfo  {journal} {Nature materials}\ }\textbf {\bibinfo
  {volume} {11}},\ \bibinfo {pages} {1023} (\bibinfo {year}
  {2012})}\BibitemShut {NoStop}%
\bibitem [{\citenamefont {Kazakov}\ \emph {et~al.}(2021)\citenamefont
  {Kazakov}, \citenamefont {Brzezicki}, \citenamefont {Hyart}, \citenamefont
  {Turowski}, \citenamefont {Polaczy\ifmmode~\acute{n}\else \'{n}\fi{}ski},
  \citenamefont {Adamus}, \citenamefont {Aleszkiewicz}, \citenamefont
  {Wojciechowski}, \citenamefont {Domagala}, \citenamefont {Caha},
  \citenamefont {Varykhalov}, \citenamefont {Springholz}, \citenamefont
  {Wojtowicz}, \citenamefont {Volobuev},\ and\ \citenamefont
  {Dietl}}]{Kazakov21}%
  \BibitemOpen
  \bibfield  {author} {\bibinfo {author} {\bibfnamefont {A.}~\bibnamefont
  {Kazakov}}, \bibinfo {author} {\bibfnamefont {W.}~\bibnamefont {Brzezicki}},
  \bibinfo {author} {\bibfnamefont {T.}~\bibnamefont {Hyart}}, \bibinfo
  {author} {\bibfnamefont {B.}~\bibnamefont {Turowski}}, \bibinfo {author}
  {\bibfnamefont {J.}~\bibnamefont {Polaczy\ifmmode~\acute{n}\else
  \'{n}\fi{}ski}}, \bibinfo {author} {\bibfnamefont {Z.}~\bibnamefont
  {Adamus}}, \bibinfo {author} {\bibfnamefont {M.}~\bibnamefont
  {Aleszkiewicz}}, \bibinfo {author} {\bibfnamefont {T.}~\bibnamefont
  {Wojciechowski}}, \bibinfo {author} {\bibfnamefont {J.~Z.}\ \bibnamefont
  {Domagala}}, \bibinfo {author} {\bibfnamefont {O.~c.~v.}\ \bibnamefont
  {Caha}}, \bibinfo {author} {\bibfnamefont {A.}~\bibnamefont {Varykhalov}},
  \bibinfo {author} {\bibfnamefont {G.}~\bibnamefont {Springholz}}, \bibinfo
  {author} {\bibfnamefont {T.}~\bibnamefont {Wojtowicz}}, \bibinfo {author}
  {\bibfnamefont {V.~V.}\ \bibnamefont {Volobuev}}, \ and\ \bibinfo {author}
  {\bibfnamefont {T.}~\bibnamefont {Dietl}},\ }\href {\doibase
  10.1103/PhysRevB.103.245307} {\bibfield  {journal} {\bibinfo  {journal}
  {Phys. Rev. B}\ }\textbf {\bibinfo {volume} {103}},\ \bibinfo {pages}
  {245307} (\bibinfo {year} {2021})}\BibitemShut {NoStop}%
\bibitem [{\citenamefont {Fu}\ \emph {et~al.}(2007)\citenamefont {Fu},
  \citenamefont {Kane},\ and\ \citenamefont {Mele}}]{Fu07}%
  \BibitemOpen
  \bibfield  {author} {\bibinfo {author} {\bibfnamefont {L.}~\bibnamefont
  {Fu}}, \bibinfo {author} {\bibfnamefont {C.~L.}\ \bibnamefont {Kane}}, \ and\
  \bibinfo {author} {\bibfnamefont {E.~J.}\ \bibnamefont {Mele}},\ }\href
  {\doibase 10.1103/PhysRevLett.98.106803} {\bibfield  {journal} {\bibinfo
  {journal} {Phys. Rev. Lett.}\ }\textbf {\bibinfo {volume} {98}},\ \bibinfo
  {pages} {106803} (\bibinfo {year} {2007})}\BibitemShut {NoStop}%
\bibitem [{\citenamefont {Teo}\ \emph {et~al.}(2008)\citenamefont {Teo},
  \citenamefont {Fu},\ and\ \citenamefont {Kane}}]{Teo08}%
  \BibitemOpen
  \bibfield  {author} {\bibinfo {author} {\bibfnamefont {J.~C.~Y.}\
  \bibnamefont {Teo}}, \bibinfo {author} {\bibfnamefont {L.}~\bibnamefont
  {Fu}}, \ and\ \bibinfo {author} {\bibfnamefont {C.~L.}\ \bibnamefont
  {Kane}},\ }\href {\doibase 10.1103/PhysRevB.78.045426} {\bibfield  {journal}
  {\bibinfo  {journal} {Phys. Rev. B}\ }\textbf {\bibinfo {volume} {78}},\
  \bibinfo {pages} {045426} (\bibinfo {year} {2008})}\BibitemShut {NoStop}%
\bibitem [{\citenamefont {Safaei}\ \emph {et~al.}(2015)\citenamefont {Safaei},
  \citenamefont {Galicka}, \citenamefont {Kacman},\ and\ \citenamefont
  {Buczko}}]{Safaei_2015}%
  \BibitemOpen
  \bibfield  {author} {\bibinfo {author} {\bibfnamefont {S.}~\bibnamefont
  {Safaei}}, \bibinfo {author} {\bibfnamefont {M.}~\bibnamefont {Galicka}},
  \bibinfo {author} {\bibfnamefont {P.}~\bibnamefont {Kacman}}, \ and\ \bibinfo
  {author} {\bibfnamefont {R.}~\bibnamefont {Buczko}},\ }\href {\doibase
  10.1088/1367-2630/17/6/063041} {\bibfield  {journal} {\bibinfo  {journal}
  {New Journal of Physics}\ }\textbf {\bibinfo {volume} {17}},\ \bibinfo
  {pages} {063041} (\bibinfo {year} {2015})}\BibitemShut {NoStop}%
\bibitem [{\citenamefont {Cuono}\ \emph {et~al.}(2022)\citenamefont {Cuono},
  \citenamefont {Hussain}, \citenamefont {Fakhredine},\ and\ \citenamefont
  {Autieri}}]{Cuono22}%
  \BibitemOpen
  \bibfield  {author} {\bibinfo {author} {\bibfnamefont {G.}~\bibnamefont
  {Cuono}}, \bibinfo {author} {\bibfnamefont {G.}~\bibnamefont {Hussain}},
  \bibinfo {author} {\bibfnamefont {A.}~\bibnamefont {Fakhredine}}, \ and\
  \bibinfo {author} {\bibfnamefont {C.}~\bibnamefont {Autieri}},\ }\href
  {\doibase 10.12693/APhysPolA.142.521} {\bibfield  {journal} {\bibinfo
  {journal} {Acta Physica Polonica A}\ }\textbf {\bibinfo {volume} {142}}
  (\bibinfo {year} {2022}),\ 10.12693/APhysPolA.142.521}\BibitemShut {NoStop}%
\bibitem [{\citenamefont {Rechci\'nski}\ \emph {et~al.}(2021)\citenamefont
  {Rechci\'nski}, \citenamefont {Galicka}, \citenamefont {Simma}, \citenamefont
  {Volobuev}, \citenamefont {Caha}, \citenamefont {Sánchez-Barriga},
  \citenamefont {Mandal}, \citenamefont {Golias}, \citenamefont {Varykhalov},
  \citenamefont {Rader}, \citenamefont {Bauer}, \citenamefont {Kacman},
  \citenamefont {Buczko},\ and\ \citenamefont {Springholz}}]{Rechcinski21}%
  \BibitemOpen
  \bibfield  {author} {\bibinfo {author} {\bibfnamefont {R.}~\bibnamefont
  {Rechci\'nski}}, \bibinfo {author} {\bibfnamefont {M.}~\bibnamefont
  {Galicka}}, \bibinfo {author} {\bibfnamefont {M.}~\bibnamefont {Simma}},
  \bibinfo {author} {\bibfnamefont {V.~V.}\ \bibnamefont {Volobuev}}, \bibinfo
  {author} {\bibfnamefont {O.}~\bibnamefont {Caha}}, \bibinfo {author}
  {\bibfnamefont {J.}~\bibnamefont {Sánchez-Barriga}}, \bibinfo {author}
  {\bibfnamefont {P.~S.}\ \bibnamefont {Mandal}}, \bibinfo {author}
  {\bibfnamefont {E.}~\bibnamefont {Golias}}, \bibinfo {author} {\bibfnamefont
  {A.}~\bibnamefont {Varykhalov}}, \bibinfo {author} {\bibfnamefont
  {O.}~\bibnamefont {Rader}}, \bibinfo {author} {\bibfnamefont
  {G.}~\bibnamefont {Bauer}}, \bibinfo {author} {\bibfnamefont
  {P.}~\bibnamefont {Kacman}}, \bibinfo {author} {\bibfnamefont
  {R.}~\bibnamefont {Buczko}}, \ and\ \bibinfo {author} {\bibfnamefont
  {G.}~\bibnamefont {Springholz}},\ }\href {\doibase
  https://doi.org/10.1002/adfm.202008885} {\bibfield  {journal} {\bibinfo
  {journal} {Advanced Functional Materials}\ }\textbf {\bibinfo {volume}
  {31}},\ \bibinfo {pages} {2008885} (\bibinfo {year} {2021})},\ \Eprint
  {http://arxiv.org/abs/https://onlinelibrary.wiley.com/doi/pdf/10.1002/adfm.202008885}
  {https://onlinelibrary.wiley.com/doi/pdf/10.1002/adfm.202008885} \BibitemShut
  {NoStop}%
\bibitem [{\citenamefont {Schindler}\ \emph {et~al.}(2018)\citenamefont
  {Schindler}, \citenamefont {Cook}, \citenamefont {Vergniory}, \citenamefont
  {Wang}, \citenamefont {Parkin}, \citenamefont {Bernevig},\ and\ \citenamefont
  {Neupert}}]{Schindler18}%
  \BibitemOpen
  \bibfield  {author} {\bibinfo {author} {\bibfnamefont {F.}~\bibnamefont
  {Schindler}}, \bibinfo {author} {\bibfnamefont {A.~M.}\ \bibnamefont {Cook}},
  \bibinfo {author} {\bibfnamefont {M.~G.}\ \bibnamefont {Vergniory}}, \bibinfo
  {author} {\bibfnamefont {Z.}~\bibnamefont {Wang}}, \bibinfo {author}
  {\bibfnamefont {S.~S.~P.}\ \bibnamefont {Parkin}}, \bibinfo {author}
  {\bibfnamefont {B.~A.}\ \bibnamefont {Bernevig}}, \ and\ \bibinfo {author}
  {\bibfnamefont {T.}~\bibnamefont {Neupert}},\ }\href {\doibase
  10.1126/sciadv.aat0346} {\bibfield  {journal} {\bibinfo  {journal} {Science
  Advances}\ }\textbf {\bibinfo {volume} {4}} (\bibinfo {year} {2018}),\
  10.1126/sciadv.aat0346}\BibitemShut {NoStop}%
\bibitem [{\citenamefont {Kooi}\ \emph {et~al.}(2020)\citenamefont {Kooi},
  \citenamefont {van Miert},\ and\ \citenamefont {Ortix}}]{Kooi20}%
  \BibitemOpen
  \bibfield  {author} {\bibinfo {author} {\bibfnamefont {S.~H.}\ \bibnamefont
  {Kooi}}, \bibinfo {author} {\bibfnamefont {G.}~\bibnamefont {van Miert}}, \
  and\ \bibinfo {author} {\bibfnamefont {C.}~\bibnamefont {Ortix}},\ }\href
  {\doibase 10.1103/PhysRevB.102.041122} {\bibfield  {journal} {\bibinfo
  {journal} {Phys. Rev. B}\ }\textbf {\bibinfo {volume} {102}},\ \bibinfo
  {pages} {041122} (\bibinfo {year} {2020})}\BibitemShut {NoStop}%
\bibitem [{\citenamefont {van Miert}\ and\ \citenamefont
  {Ortix}(2018)}]{vanMiert18}%
  \BibitemOpen
  \bibfield  {author} {\bibinfo {author} {\bibfnamefont {G.}~\bibnamefont {van
  Miert}}\ and\ \bibinfo {author} {\bibfnamefont {C.}~\bibnamefont {Ortix}},\
  }\href {\doibase 10.1103/PhysRevB.98.081110} {\bibfield  {journal} {\bibinfo
  {journal} {Phys. Rev. B}\ }\textbf {\bibinfo {volume} {98}},\ \bibinfo
  {pages} {081110} (\bibinfo {year} {2018})}\BibitemShut {NoStop}%
\bibitem [{\citenamefont {Plekhanov}\ \emph {et~al.}(2014)\citenamefont
  {Plekhanov}, \citenamefont {Barone}, \citenamefont {Di~Sante},\ and\
  \citenamefont {Picozzi}}]{Plekhanov14}%
  \BibitemOpen
  \bibfield  {author} {\bibinfo {author} {\bibfnamefont {E.}~\bibnamefont
  {Plekhanov}}, \bibinfo {author} {\bibfnamefont {P.}~\bibnamefont {Barone}},
  \bibinfo {author} {\bibfnamefont {D.}~\bibnamefont {Di~Sante}}, \ and\
  \bibinfo {author} {\bibfnamefont {S.}~\bibnamefont {Picozzi}},\ }\href
  {\doibase 10.1103/PhysRevB.90.161108} {\bibfield  {journal} {\bibinfo
  {journal} {Phys. Rev. B}\ }\textbf {\bibinfo {volume} {90}},\ \bibinfo
  {pages} {161108} (\bibinfo {year} {2014})}\BibitemShut {NoStop}%
\bibitem [{\citenamefont {Wang}\ \emph {et~al.}(2020)\citenamefont {Wang},
  \citenamefont {Gopal}, \citenamefont {Picozzi}, \citenamefont {Curtarolo},
  \citenamefont {Buongiorno~Nardelli},\ and\ \citenamefont
  {Sławińska}}]{Wang20}%
  \BibitemOpen
  \bibfield  {author} {\bibinfo {author} {\bibfnamefont {H.}~\bibnamefont
  {Wang}}, \bibinfo {author} {\bibfnamefont {P.}~\bibnamefont {Gopal}},
  \bibinfo {author} {\bibfnamefont {S.}~\bibnamefont {Picozzi}}, \bibinfo
  {author} {\bibfnamefont {S.}~\bibnamefont {Curtarolo}}, \bibinfo {author}
  {\bibfnamefont {M.}~\bibnamefont {Buongiorno~Nardelli}}, \ and\ \bibinfo
  {author} {\bibfnamefont {J.}~\bibnamefont {Sławińska}},\ }\href {\doibase
  10.1038/s41524-020-0274-0} {\bibfield  {journal} {\bibinfo  {journal} {npj
  Computational Materials}\ }\textbf {\bibinfo {volume} {6}},\ \bibinfo {pages}
  {161108} (\bibinfo {year} {2020})}\BibitemShut {NoStop}%
\bibitem [{\citenamefont {Barone}\ \emph {et~al.}(2013)\citenamefont {Barone},
  \citenamefont {Rauch}, \citenamefont {Di~Sante}, \citenamefont {Henk},
  \citenamefont {Mertig},\ and\ \citenamefont {Picozzi}}]{Barone13}%
  \BibitemOpen
  \bibfield  {author} {\bibinfo {author} {\bibfnamefont {P.}~\bibnamefont
  {Barone}}, \bibinfo {author} {\bibfnamefont {T.~c.~v.}\ \bibnamefont
  {Rauch}}, \bibinfo {author} {\bibfnamefont {D.}~\bibnamefont {Di~Sante}},
  \bibinfo {author} {\bibfnamefont {J.}~\bibnamefont {Henk}}, \bibinfo {author}
  {\bibfnamefont {I.}~\bibnamefont {Mertig}}, \ and\ \bibinfo {author}
  {\bibfnamefont {S.}~\bibnamefont {Picozzi}},\ }\href {\doibase
  10.1103/PhysRevB.88.045207} {\bibfield  {journal} {\bibinfo  {journal} {Phys.
  Rev. B}\ }\textbf {\bibinfo {volume} {88}},\ \bibinfo {pages} {045207}
  (\bibinfo {year} {2013})}\BibitemShut {NoStop}%
\bibitem [{\citenamefont {S{\l}awi{\'{n}}ska}\ \emph
  {et~al.}(2020)\citenamefont {S{\l}awi{\'{n}}ska}, \citenamefont {Cerasoli},
  \citenamefont {Gopal}, \citenamefont {Costa}, \citenamefont {Curtarolo},\
  and\ \citenamefont {Nardelli}}]{Slawinska20}%
  \BibitemOpen
  \bibfield  {author} {\bibinfo {author} {\bibfnamefont {J.}~\bibnamefont
  {S{\l}awi{\'{n}}ska}}, \bibinfo {author} {\bibfnamefont {F.~T.}\ \bibnamefont
  {Cerasoli}}, \bibinfo {author} {\bibfnamefont {P.}~\bibnamefont {Gopal}},
  \bibinfo {author} {\bibfnamefont {M.}~\bibnamefont {Costa}}, \bibinfo
  {author} {\bibfnamefont {S.}~\bibnamefont {Curtarolo}}, \ and\ \bibinfo
  {author} {\bibfnamefont {M.~B.}\ \bibnamefont {Nardelli}},\ }\href {\doibase
  10.1088/2053-1583/ab6f7a} {\bibfield  {journal} {\bibinfo  {journal} {2D
  Materials}\ }\textbf {\bibinfo {volume} {7}},\ \bibinfo {pages} {025026}
  (\bibinfo {year} {2020})}\BibitemShut {NoStop}%
\bibitem [{\citenamefont {Volobuev}\ \emph {et~al.}(2017)\citenamefont
  {Volobuev}, \citenamefont {Mandal}, \citenamefont {Galicka}, \citenamefont
  {Caha}, \citenamefont {Sánchez-Barriga}, \citenamefont {Di~Sante},
  \citenamefont {Varykhalov}, \citenamefont {Khiar}, \citenamefont {Picozzi},
  \citenamefont {Bauer}, \citenamefont {Kacman}, \citenamefont {Buczko},
  \citenamefont {Rader},\ and\ \citenamefont {Springholz}}]{Volobuev17}%
  \BibitemOpen
  \bibfield  {author} {\bibinfo {author} {\bibfnamefont {V.~V.}\ \bibnamefont
  {Volobuev}}, \bibinfo {author} {\bibfnamefont {P.~S.}\ \bibnamefont
  {Mandal}}, \bibinfo {author} {\bibfnamefont {M.}~\bibnamefont {Galicka}},
  \bibinfo {author} {\bibfnamefont {O.}~\bibnamefont {Caha}}, \bibinfo {author}
  {\bibfnamefont {J.}~\bibnamefont {Sánchez-Barriga}}, \bibinfo {author}
  {\bibfnamefont {D.}~\bibnamefont {Di~Sante}}, \bibinfo {author}
  {\bibfnamefont {A.}~\bibnamefont {Varykhalov}}, \bibinfo {author}
  {\bibfnamefont {A.}~\bibnamefont {Khiar}}, \bibinfo {author} {\bibfnamefont
  {S.}~\bibnamefont {Picozzi}}, \bibinfo {author} {\bibfnamefont
  {G.}~\bibnamefont {Bauer}}, \bibinfo {author} {\bibfnamefont
  {P.}~\bibnamefont {Kacman}}, \bibinfo {author} {\bibfnamefont
  {R.}~\bibnamefont {Buczko}}, \bibinfo {author} {\bibfnamefont
  {O.}~\bibnamefont {Rader}}, \ and\ \bibinfo {author} {\bibfnamefont
  {G.}~\bibnamefont {Springholz}},\ }\href {\doibase
  https://doi.org/10.1002/adma.201604185} {\bibfield  {journal} {\bibinfo
  {journal} {Advanced Materials}\ }\textbf {\bibinfo {volume} {29}},\ \bibinfo
  {pages} {1604185} (\bibinfo {year} {2017})}\BibitemShut {NoStop}%
\bibitem [{\citenamefont {Liu}\ \emph {et~al.}(2018)\citenamefont {Liu},
  \citenamefont {Lu}, \citenamefont {Picozzi}, \citenamefont {Bellaiche},\ and\
  \citenamefont {Xiang}}]{Liu18}%
  \BibitemOpen
  \bibfield  {author} {\bibinfo {author} {\bibfnamefont {K.}~\bibnamefont
  {Liu}}, \bibinfo {author} {\bibfnamefont {J.}~\bibnamefont {Lu}}, \bibinfo
  {author} {\bibfnamefont {S.}~\bibnamefont {Picozzi}}, \bibinfo {author}
  {\bibfnamefont {L.}~\bibnamefont {Bellaiche}}, \ and\ \bibinfo {author}
  {\bibfnamefont {H.}~\bibnamefont {Xiang}},\ }\href {\doibase
  10.1103/PhysRevLett.121.027601} {\bibfield  {journal} {\bibinfo  {journal}
  {Phys. Rev. Lett.}\ }\textbf {\bibinfo {volume} {121}},\ \bibinfo {pages}
  {027601} (\bibinfo {year} {2018})}\BibitemShut {NoStop}%
\bibitem [{\citenamefont {Liu}\ \emph {et~al.}(2014)\citenamefont {Liu},
  \citenamefont {Hsieh}, \citenamefont {Wei}, \citenamefont {Duan},
  \citenamefont {Moodera},\ and\ \citenamefont {Fu}}]{liu2014spin}%
  \BibitemOpen
  \bibfield  {author} {\bibinfo {author} {\bibfnamefont {J.}~\bibnamefont
  {Liu}}, \bibinfo {author} {\bibfnamefont {T.~H.}\ \bibnamefont {Hsieh}},
  \bibinfo {author} {\bibfnamefont {P.}~\bibnamefont {Wei}}, \bibinfo {author}
  {\bibfnamefont {W.}~\bibnamefont {Duan}}, \bibinfo {author} {\bibfnamefont
  {J.}~\bibnamefont {Moodera}}, \ and\ \bibinfo {author} {\bibfnamefont
  {L.}~\bibnamefont {Fu}},\ }\href@noop {} {\bibfield  {journal} {\bibinfo
  {journal} {Nature materials}\ }\textbf {\bibinfo {volume} {13}},\ \bibinfo
  {pages} {178} (\bibinfo {year} {2014})}\BibitemShut {NoStop}%
\bibitem [{\citenamefont {Liu}\ \emph {et~al.}(2015)\citenamefont {Liu},
  \citenamefont {Qian},\ and\ \citenamefont {Fu}}]{liu2015crystal}%
  \BibitemOpen
  \bibfield  {author} {\bibinfo {author} {\bibfnamefont {J.}~\bibnamefont
  {Liu}}, \bibinfo {author} {\bibfnamefont {X.}~\bibnamefont {Qian}}, \ and\
  \bibinfo {author} {\bibfnamefont {L.}~\bibnamefont {Fu}},\ }\href@noop {}
  {\bibfield  {journal} {\bibinfo  {journal} {Nano letters}\ }\textbf {\bibinfo
  {volume} {15}},\ \bibinfo {pages} {2657} (\bibinfo {year}
  {2015})}\BibitemShut {NoStop}%
\bibitem [{\citenamefont {Arroyo-Gascón}\ \emph {et~al.}(2022)\citenamefont
  {Arroyo-Gascón}, \citenamefont {Baba}, \citenamefont {Cerdá}, \citenamefont
  {de~Abril}, \citenamefont {Martínez}, \citenamefont {Domínguez-Adame},\
  and\ \citenamefont {Chico}}]{D1NR07120C}%
  \BibitemOpen
  \bibfield  {author} {\bibinfo {author} {\bibfnamefont {O.}~\bibnamefont
  {Arroyo-Gascón}}, \bibinfo {author} {\bibfnamefont {Y.}~\bibnamefont
  {Baba}}, \bibinfo {author} {\bibfnamefont {J.~I.}\ \bibnamefont {Cerdá}},
  \bibinfo {author} {\bibfnamefont {O.}~\bibnamefont {de~Abril}}, \bibinfo
  {author} {\bibfnamefont {R.}~\bibnamefont {Martínez}}, \bibinfo {author}
  {\bibfnamefont {F.}~\bibnamefont {Domínguez-Adame}}, \ and\ \bibinfo
  {author} {\bibfnamefont {L.}~\bibnamefont {Chico}},\ }\href {\doibase
  10.1039/D1NR07120C} {\bibfield  {journal} {\bibinfo  {journal} {Nanoscale}\
  }\textbf {\bibinfo {volume} {14}},\ \bibinfo {pages} {7151} (\bibinfo {year}
  {2022})}\BibitemShut {NoStop}%
\bibitem [{\citenamefont {Samadi}\ \emph {et~al.}(2023)\citenamefont {Samadi},
  \citenamefont {Rechci\ifmmode~\acute{n}\else \'{n}\fi{}ski},\ and\
  \citenamefont {Buczko}}]{Samadi23}%
  \BibitemOpen
  \bibfield  {author} {\bibinfo {author} {\bibfnamefont {S.}~\bibnamefont
  {Samadi}}, \bibinfo {author} {\bibfnamefont {R.}~\bibnamefont
  {Rechci\ifmmode~\acute{n}\else \'{n}\fi{}ski}}, \ and\ \bibinfo {author}
  {\bibfnamefont {R.}~\bibnamefont {Buczko}},\ }\href {\doibase
  10.1103/PhysRevB.107.205401} {\bibfield  {journal} {\bibinfo  {journal}
  {Phys. Rev. B}\ }\textbf {\bibinfo {volume} {107}},\ \bibinfo {pages}
  {205401} (\bibinfo {year} {2023})}\BibitemShut {NoStop}%
\bibitem [{\citenamefont {Sessi}\ \emph {et~al.}(2016)\citenamefont {Sessi},
  \citenamefont {Di~Sante}, \citenamefont {Szczerbakow}, \citenamefont {Glott},
  \citenamefont {Wilfert}, \citenamefont {Shmidt}, \citenamefont {Bathon},
  \citenamefont {Dziawa}, \citenamefont {Greiter}, \citenamefont {Neupert},
  \citenamefont {Greiter}, \citenamefont {Neupert}, \citenamefont
  {Sangiovanni}, \citenamefont {Story}, \citenamefont {Thomale},\ and\
  \citenamefont {Bode}}]{Sessi16}%
  \BibitemOpen
  \bibfield  {author} {\bibinfo {author} {\bibfnamefont {P.}~\bibnamefont
  {Sessi}}, \bibinfo {author} {\bibfnamefont {D.}~\bibnamefont {Di~Sante}},
  \bibinfo {author} {\bibfnamefont {A.}~\bibnamefont {Szczerbakow}}, \bibinfo
  {author} {\bibfnamefont {F.}~\bibnamefont {Glott}}, \bibinfo {author}
  {\bibfnamefont {S.}~\bibnamefont {Wilfert}}, \bibinfo {author} {\bibfnamefont
  {H.}~\bibnamefont {Shmidt}}, \bibinfo {author} {\bibfnamefont
  {T.}~\bibnamefont {Bathon}}, \bibinfo {author} {\bibfnamefont
  {P.}~\bibnamefont {Dziawa}}, \bibinfo {author} {\bibfnamefont
  {M.}~\bibnamefont {Greiter}}, \bibinfo {author} {\bibfnamefont
  {M.}~\bibnamefont {Neupert}}, \bibinfo {author} {\bibfnamefont
  {M.}~\bibnamefont {Greiter}}, \bibinfo {author} {\bibfnamefont
  {T.}~\bibnamefont {Neupert}}, \bibinfo {author} {\bibfnamefont
  {G.}~\bibnamefont {Sangiovanni}}, \bibinfo {author} {\bibfnamefont
  {T.}~\bibnamefont {Story}}, \bibinfo {author} {\bibfnamefont
  {R.}~\bibnamefont {Thomale}}, \ and\ \bibinfo {author} {\bibfnamefont
  {M.}~\bibnamefont {Bode}},\ }\href {\doibase 10.1126/science.aah6233}
  {\bibfield  {journal} {\bibinfo  {journal} {Science}\ }\textbf {\bibinfo
  {volume} {354}},\ \bibinfo {pages} {1269} (\bibinfo {year}
  {2016})}\BibitemShut {NoStop}%
\bibitem [{\citenamefont {Brzezicki}\ \emph {et~al.}(2019)\citenamefont
  {Brzezicki}, \citenamefont {Wysoki\ifmmode~\acute{n}\else \'{n}\fi{}ski},\
  and\ \citenamefont {Hyart}}]{Brzezicki19}%
  \BibitemOpen
  \bibfield  {author} {\bibinfo {author} {\bibfnamefont {W.}~\bibnamefont
  {Brzezicki}}, \bibinfo {author} {\bibfnamefont {M.~M.}\ \bibnamefont
  {Wysoki\ifmmode~\acute{n}\else \'{n}\fi{}ski}}, \ and\ \bibinfo {author}
  {\bibfnamefont {T.}~\bibnamefont {Hyart}},\ }\href {\doibase
  10.1103/PhysRevB.100.121107} {\bibfield  {journal} {\bibinfo  {journal}
  {Phys. Rev. B}\ }\textbf {\bibinfo {volume} {100}},\ \bibinfo {pages}
  {121107} (\bibinfo {year} {2019})}\BibitemShut {NoStop}%
\bibitem [{\citenamefont {Wagner}\ \emph {et~al.}(2012)\citenamefont {Wagner},
  \citenamefont {Das}, \citenamefont {Jung}, \citenamefont {Odobesko},
  \citenamefont {Küster}, \citenamefont {Keller}, \citenamefont {Korczak},
  \citenamefont {Szczerbakow}, \citenamefont {Story}, \citenamefont {Parkin},
  \citenamefont {Thomale}, \citenamefont {Neupert}, \citenamefont {Bode},\ and\
  \citenamefont {Sessi}}]{Wagner12}%
  \BibitemOpen
  \bibfield  {author} {\bibinfo {author} {\bibfnamefont {G.}~\bibnamefont
  {Wagner}}, \bibinfo {author} {\bibfnamefont {S.}~\bibnamefont {Das}},
  \bibinfo {author} {\bibfnamefont {J.}~\bibnamefont {Jung}}, \bibinfo {author}
  {\bibfnamefont {A.}~\bibnamefont {Odobesko}}, \bibinfo {author}
  {\bibfnamefont {F.}~\bibnamefont {Küster}}, \bibinfo {author} {\bibfnamefont
  {F.}~\bibnamefont {Keller}}, \bibinfo {author} {\bibfnamefont
  {J.}~\bibnamefont {Korczak}}, \bibinfo {author} {\bibfnamefont
  {A.}~\bibnamefont {Szczerbakow}}, \bibinfo {author} {\bibfnamefont
  {T.}~\bibnamefont {Story}}, \bibinfo {author} {\bibfnamefont {S.~S.~P.}\
  \bibnamefont {Parkin}}, \bibinfo {author} {\bibfnamefont {R.}~\bibnamefont
  {Thomale}}, \bibinfo {author} {\bibfnamefont {T.}~\bibnamefont {Neupert}},
  \bibinfo {author} {\bibfnamefont {M.}~\bibnamefont {Bode}}, \ and\ \bibinfo
  {author} {\bibfnamefont {P.}~\bibnamefont {Sessi}},\ }\href {\doibase
  10.1021/acs.nanolett.2c03794} {\bibfield  {journal} {\bibinfo  {journal}
  {Nano Letters}\ }\textbf {\bibinfo {volume} {23}},\ \bibinfo {pages} {2476}
  (\bibinfo {year} {2012})}\BibitemShut {NoStop}%
\bibitem [{\citenamefont {Shen}\ and\ \citenamefont {Cha}(2014)}]{C4NR05124F}%
  \BibitemOpen
  \bibfield  {author} {\bibinfo {author} {\bibfnamefont {J.}~\bibnamefont
  {Shen}}\ and\ \bibinfo {author} {\bibfnamefont {J.~J.}\ \bibnamefont {Cha}},\
  }\href {\doibase 10.1039/C4NR05124F} {\bibfield  {journal} {\bibinfo
  {journal} {Nanoscale}\ }\textbf {\bibinfo {volume} {6}},\ \bibinfo {pages}
  {14133} (\bibinfo {year} {2014})}\BibitemShut {NoStop}%
\bibitem [{\citenamefont {Sadowski}\ \emph {et~al.}(2018)\citenamefont
  {Sadowski}, \citenamefont {Dziawa}, \citenamefont {Kaleta}, \citenamefont
  {Kurowska}, \citenamefont {Reszka}, \citenamefont {Story},\ and\
  \citenamefont {Kret}}]{Sadowski2018}%
  \BibitemOpen
  \bibfield  {author} {\bibinfo {author} {\bibfnamefont {J.}~\bibnamefont
  {Sadowski}}, \bibinfo {author} {\bibfnamefont {P.}~\bibnamefont {Dziawa}},
  \bibinfo {author} {\bibfnamefont {A.}~\bibnamefont {Kaleta}}, \bibinfo
  {author} {\bibfnamefont {B.}~\bibnamefont {Kurowska}}, \bibinfo {author}
  {\bibfnamefont {A.}~\bibnamefont {Reszka}}, \bibinfo {author} {\bibfnamefont
  {T.}~\bibnamefont {Story}}, \ and\ \bibinfo {author} {\bibfnamefont
  {S.}~\bibnamefont {Kret}},\ }\href {\doibase 10.1039/C8NR06096G} {\bibfield
  {journal} {\bibinfo  {journal} {Nanoscale}\ }\textbf {\bibinfo {volume}
  {10}},\ \bibinfo {pages} {20772} (\bibinfo {year} {2018})}\BibitemShut
  {NoStop}%
\bibitem [{\citenamefont {Vasylenko}\ \emph {et~al.}(2018)\citenamefont
  {Vasylenko}, \citenamefont {Marks}, \citenamefont {Wynn}, \citenamefont
  {Medeiros}, \citenamefont {Ramasse}, \citenamefont {Morris}, \citenamefont
  {Sloan},\ and\ \citenamefont {Quigley}}]{Vasylenko18}%
  \BibitemOpen
  \bibfield  {author} {\bibinfo {author} {\bibfnamefont {A.}~\bibnamefont
  {Vasylenko}}, \bibinfo {author} {\bibfnamefont {S.}~\bibnamefont {Marks}},
  \bibinfo {author} {\bibfnamefont {J.~M.}\ \bibnamefont {Wynn}}, \bibinfo
  {author} {\bibfnamefont {P.~V.~C.}\ \bibnamefont {Medeiros}}, \bibinfo
  {author} {\bibfnamefont {Q.~M.}\ \bibnamefont {Ramasse}}, \bibinfo {author}
  {\bibfnamefont {A.~J.}\ \bibnamefont {Morris}}, \bibinfo {author}
  {\bibfnamefont {J.}~\bibnamefont {Sloan}}, \ and\ \bibinfo {author}
  {\bibfnamefont {D.}~\bibnamefont {Quigley}},\ }\href {\doibase
  10.1021/acsnano.8b02261} {\bibfield  {journal} {\bibinfo  {journal} {ACS
  Nano}\ }\textbf {\bibinfo {volume} {12}},\ \bibinfo {pages} {6023} (\bibinfo
  {year} {2018})}\BibitemShut {NoStop}%
\bibitem [{\citenamefont {Skiff}\ \emph
  {et~al.}(2023{\natexlab{a}})\citenamefont {Skiff}, \citenamefont {de~Juan},
  \citenamefont {Queiroz}, \citenamefont {Mathimalar}, \citenamefont
  {Beidenkopf},\ and\ \citenamefont {Ilan}}]{Skiff22}%
  \BibitemOpen
  \bibfield  {author} {\bibinfo {author} {\bibfnamefont {R.~M.}\ \bibnamefont
  {Skiff}}, \bibinfo {author} {\bibfnamefont {F.}~\bibnamefont {de~Juan}},
  \bibinfo {author} {\bibfnamefont {R.}~\bibnamefont {Queiroz}}, \bibinfo
  {author} {\bibfnamefont {S.}~\bibnamefont {Mathimalar}}, \bibinfo {author}
  {\bibfnamefont {H.}~\bibnamefont {Beidenkopf}}, \ and\ \bibinfo {author}
  {\bibfnamefont {R.}~\bibnamefont {Ilan}},\ }\href {\doibase
  10.21468/SciPostPhysCore.6.1.011} {\bibfield  {journal} {\bibinfo  {journal}
  {SciPost Phys. Core}\ }\textbf {\bibinfo {volume} {6}},\ \bibinfo {pages}
  {011} (\bibinfo {year} {2023}{\natexlab{a}})}\BibitemShut {NoStop}%
\bibitem [{\citenamefont {Safdar}\ \emph {et~al.}(2015)\citenamefont {Safdar},
  \citenamefont {Wang}, \citenamefont {Wang}, \citenamefont {Zhan},
  \citenamefont {Xu}, \citenamefont {Wang}, \citenamefont {Mirza},\ and\
  \citenamefont {He}}]{Safdar15}%
  \BibitemOpen
  \bibfield  {author} {\bibinfo {author} {\bibfnamefont {M.}~\bibnamefont
  {Safdar}}, \bibinfo {author} {\bibfnamefont {Q.}~\bibnamefont {Wang}},
  \bibinfo {author} {\bibfnamefont {Z.}~\bibnamefont {Wang}}, \bibinfo {author}
  {\bibfnamefont {X.}~\bibnamefont {Zhan}}, \bibinfo {author} {\bibfnamefont
  {K.}~\bibnamefont {Xu}}, \bibinfo {author} {\bibfnamefont {F.}~\bibnamefont
  {Wang}}, \bibinfo {author} {\bibfnamefont {M.}~\bibnamefont {Mirza}}, \ and\
  \bibinfo {author} {\bibfnamefont {J.}~\bibnamefont {He}},\ }\href {\doibase
  10.1021/nl504976g} {\bibfield  {journal} {\bibinfo  {journal} {Nano Letters}\
  }\textbf {\bibinfo {volume} {15}} (\bibinfo {year} {2015}),\
  10.1021/nl504976g}\BibitemShut {NoStop}%
\bibitem [{\citenamefont {Dad}\ \emph {et~al.}(2022)\citenamefont {Dad},
  \citenamefont {Dziawa}, \citenamefont {Zajkowska}, \citenamefont {Kret},
  \citenamefont {Kozłowski}, \citenamefont {Wójcik},\ and\ \citenamefont
  {Sadowski}}]{dad2022nearly}%
  \BibitemOpen
  \bibfield  {author} {\bibinfo {author} {\bibfnamefont {S.}~\bibnamefont
  {Dad}}, \bibinfo {author} {\bibfnamefont {P.}~\bibnamefont {Dziawa}},
  \bibinfo {author} {\bibfnamefont {W.}~\bibnamefont {Zajkowska}}, \bibinfo
  {author} {\bibfnamefont {S.}~\bibnamefont {Kret}}, \bibinfo {author}
  {\bibfnamefont {M.}~\bibnamefont {Kozłowski}}, \bibinfo {author}
  {\bibfnamefont {M.}~\bibnamefont {Wójcik}}, \ and\ \bibinfo {author}
  {\bibfnamefont {J.}~\bibnamefont {Sadowski}},\ }\href@noop {} {\enquote
  {\bibinfo {title} {Nearly lattice matched gaas/pb(1-x)sn(x)te core-shell
  nanowires},}\ } (\bibinfo {year} {2022}),\ \Eprint
  {http://arxiv.org/abs/2211.08154} {arXiv:2211.08154 [cond-mat.mtrl-sci]}
  \BibitemShut {NoStop}%
\bibitem [{\citenamefont {Saghir}\ \emph {et~al.}(2015)\citenamefont {Saghir},
  \citenamefont {Sanchez}, \citenamefont {Hindmarsh}, \citenamefont {York},\
  and\ \citenamefont {Balakrishnan}}]{Saghir15}%
  \BibitemOpen
  \bibfield  {author} {\bibinfo {author} {\bibfnamefont {M.}~\bibnamefont
  {Saghir}}, \bibinfo {author} {\bibfnamefont {A.~M.}\ \bibnamefont {Sanchez}},
  \bibinfo {author} {\bibfnamefont {S.~A.}\ \bibnamefont {Hindmarsh}}, \bibinfo
  {author} {\bibfnamefont {S.~J.}\ \bibnamefont {York}}, \ and\ \bibinfo
  {author} {\bibfnamefont {G.}~\bibnamefont {Balakrishnan}},\ }\href {\doibase
  10.1021/acs.cgd.5b00577} {\bibfield  {journal} {\bibinfo  {journal} {Cryst.
  Growth Des.}\ }\textbf {\bibinfo {volume} {15}} (\bibinfo {year} {2015}),\
  10.1021/acs.cgd.5b00577}\BibitemShut {NoStop}%
\bibitem [{\citenamefont {Xu}\ \emph {et~al.}(2016)\citenamefont {Xu},
  \citenamefont {Li}, \citenamefont {Acosta}, \citenamefont {Li}, \citenamefont
  {Swartzentruber}, \citenamefont {Zheng}, \citenamefont {Sinitsyn},
  \citenamefont {Htoon}, \citenamefont {Wang},\ and\ \citenamefont
  {Zhang}}]{Xu2016-vw}%
  \BibitemOpen
  \bibfield  {author} {\bibinfo {author} {\bibfnamefont {E.}~\bibnamefont
  {Xu}}, \bibinfo {author} {\bibfnamefont {Z.}~\bibnamefont {Li}}, \bibinfo
  {author} {\bibfnamefont {J.~A.}\ \bibnamefont {Acosta}}, \bibinfo {author}
  {\bibfnamefont {N.}~\bibnamefont {Li}}, \bibinfo {author} {\bibfnamefont
  {B.}~\bibnamefont {Swartzentruber}}, \bibinfo {author} {\bibfnamefont
  {S.}~\bibnamefont {Zheng}}, \bibinfo {author} {\bibfnamefont
  {N.}~\bibnamefont {Sinitsyn}}, \bibinfo {author} {\bibfnamefont
  {H.}~\bibnamefont {Htoon}}, \bibinfo {author} {\bibfnamefont
  {J.}~\bibnamefont {Wang}}, \ and\ \bibinfo {author} {\bibfnamefont
  {S.}~\bibnamefont {Zhang}},\ }\href@noop {} {\bibfield  {journal} {\bibinfo
  {journal} {Nano Res.}\ }\textbf {\bibinfo {volume} {9}},\ \bibinfo {pages}
  {820} (\bibinfo {year} {2016})}\BibitemShut {NoStop}%
\bibitem [{\citenamefont {Hussain}\ \emph {et~al.}(2023)\citenamefont
  {Hussain}, \citenamefont {Cuono}, \citenamefont {et~al.},\ and\ \citenamefont
  {Autieri}}]{Hussain23pentagonal}%
  \BibitemOpen
  \bibfield  {author} {\bibinfo {author} {\bibfnamefont {G.}~\bibnamefont
  {Hussain}}, \bibinfo {author} {\bibfnamefont {G.}~\bibnamefont {Cuono}},
  \bibinfo {author} {\bibnamefont {et~al.}}, \ and\ \bibinfo {author}
  {\bibfnamefont {C.}~\bibnamefont {Autieri}},\ }\href@noop {} {\bibfield
  {journal} {\bibinfo  {journal} {In manuscript}\ } (\bibinfo {year}
  {2023})}\BibitemShut {NoStop}%
\bibitem [{\citenamefont {Nguyen}\ \emph {et~al.}(2022)\citenamefont {Nguyen},
  \citenamefont {Brzezicki},\ and\ \citenamefont {Hyart}}]{Nguyen22}%
  \BibitemOpen
  \bibfield  {author} {\bibinfo {author} {\bibfnamefont {N.~M.}\ \bibnamefont
  {Nguyen}}, \bibinfo {author} {\bibfnamefont {W.}~\bibnamefont {Brzezicki}}, \
  and\ \bibinfo {author} {\bibfnamefont {T.}~\bibnamefont {Hyart}},\ }\href
  {\doibase 10.1103/PhysRevB.105.075310} {\bibfield  {journal} {\bibinfo
  {journal} {Phys. Rev. B}\ }\textbf {\bibinfo {volume} {105}},\ \bibinfo
  {pages} {075310} (\bibinfo {year} {2022})}\BibitemShut {NoStop}%
\bibitem [{\citenamefont {Shukla}\ \emph {et~al.}(2022)\citenamefont {Shukla},
  \citenamefont {Zala}, \citenamefont {Gupta},\ and\ \citenamefont
  {Gajjar}}]{Shukla22}%
  \BibitemOpen
  \bibfield  {author} {\bibinfo {author} {\bibfnamefont {R.~S.}\ \bibnamefont
  {Shukla}}, \bibinfo {author} {\bibfnamefont {V.~B.}\ \bibnamefont {Zala}},
  \bibinfo {author} {\bibfnamefont {S.~K.}\ \bibnamefont {Gupta}}, \ and\
  \bibinfo {author} {\bibfnamefont {P.~N.}\ \bibnamefont {Gajjar}},\ }\href
  {\doibase 10.1039/D2TC03400J} {\bibfield  {journal} {\bibinfo  {journal} {J.
  Mater. Chem. C}\ }\textbf {\bibinfo {volume} {10}},\ \bibinfo {pages} {15601}
  (\bibinfo {year} {2022})}\BibitemShut {NoStop}%
\bibitem [{\citenamefont {Song}\ \emph {et~al.}(2023)\citenamefont {Song},
  \citenamefont {Wang}, \citenamefont {Miao}, \citenamefont {Yu}, \citenamefont
  {Gao}, \citenamefont {Li}, \citenamefont {Yang}, \citenamefont {Chen},
  \citenamefont {Geng}, \citenamefont {Zhang}, \citenamefont {Zhang},
  \citenamefont {Zang}, \citenamefont {Cao}, \citenamefont {Liu}, \citenamefont
  {Shang}, \citenamefont {Feng}, \citenamefont {Li}, \citenamefont {Xue},
  \citenamefont {He},\ and\ \citenamefont {Zhang}}]{Song23}%
  \BibitemOpen
  \bibfield  {author} {\bibinfo {author} {\bibfnamefont {W.}~\bibnamefont
  {Song}}, \bibinfo {author} {\bibfnamefont {Y.}~\bibnamefont {Wang}}, \bibinfo
  {author} {\bibfnamefont {W.}~\bibnamefont {Miao}}, \bibinfo {author}
  {\bibfnamefont {Z.}~\bibnamefont {Yu}}, \bibinfo {author} {\bibfnamefont
  {Y.}~\bibnamefont {Gao}}, \bibinfo {author} {\bibfnamefont {R.}~\bibnamefont
  {Li}}, \bibinfo {author} {\bibfnamefont {S.}~\bibnamefont {Yang}}, \bibinfo
  {author} {\bibfnamefont {F.}~\bibnamefont {Chen}}, \bibinfo {author}
  {\bibfnamefont {Z.}~\bibnamefont {Geng}}, \bibinfo {author} {\bibfnamefont
  {Z.}~\bibnamefont {Zhang}}, \bibinfo {author} {\bibfnamefont
  {S.}~\bibnamefont {Zhang}}, \bibinfo {author} {\bibfnamefont
  {Y.}~\bibnamefont {Zang}}, \bibinfo {author} {\bibfnamefont {Z.}~\bibnamefont
  {Cao}}, \bibinfo {author} {\bibfnamefont {D.~E.}\ \bibnamefont {Liu}},
  \bibinfo {author} {\bibfnamefont {R.}~\bibnamefont {Shang}}, \bibinfo
  {author} {\bibfnamefont {X.}~\bibnamefont {Feng}}, \bibinfo {author}
  {\bibfnamefont {L.}~\bibnamefont {Li}}, \bibinfo {author} {\bibfnamefont
  {Q.-K.}\ \bibnamefont {Xue}}, \bibinfo {author} {\bibfnamefont
  {K.}~\bibnamefont {He}}, \ and\ \bibinfo {author} {\bibfnamefont
  {H.}~\bibnamefont {Zhang}},\ }\href@noop {} {\enquote {\bibinfo {title}
  {Conductance quantization in pbte nanowires},}\ } (\bibinfo {year} {2023}),\
  \Eprint {http://arxiv.org/abs/2304.10194} {arXiv:2304.10194
  [cond-mat.mes-hall]} \BibitemShut {NoStop}%
\bibitem [{\citenamefont {Jung}\ \emph {et~al.}(2022)\citenamefont {Jung},
  \citenamefont {Schellingerhout}, \citenamefont {Ritter}, \citenamefont {ten
  Kate}, \citenamefont {van~der Molen}, \citenamefont {de~Loijer},
  \citenamefont {Verheijen}, \citenamefont {Riel}, \citenamefont {Nichele},\
  and\ \citenamefont {Bakkers}}]{Jung22}%
  \BibitemOpen
  \bibfield  {author} {\bibinfo {author} {\bibfnamefont {J.}~\bibnamefont
  {Jung}}, \bibinfo {author} {\bibfnamefont {S.~G.}\ \bibnamefont
  {Schellingerhout}}, \bibinfo {author} {\bibfnamefont {M.~F.}\ \bibnamefont
  {Ritter}}, \bibinfo {author} {\bibfnamefont {S.~C.}\ \bibnamefont {ten
  Kate}}, \bibinfo {author} {\bibfnamefont {O.~A.}\ \bibnamefont {van~der
  Molen}}, \bibinfo {author} {\bibfnamefont {S.}~\bibnamefont {de~Loijer}},
  \bibinfo {author} {\bibfnamefont {M.~A.}\ \bibnamefont {Verheijen}}, \bibinfo
  {author} {\bibfnamefont {H.}~\bibnamefont {Riel}}, \bibinfo {author}
  {\bibfnamefont {F.}~\bibnamefont {Nichele}}, \ and\ \bibinfo {author}
  {\bibfnamefont {E.~P.}\ \bibnamefont {Bakkers}},\ }\href {\doibase
  10.1002/adfm.202208974} {\bibfield  {journal} {\bibinfo  {journal} {Advanced
  Functional Materials}\ }\textbf {\bibinfo {volume} {32}} (\bibinfo {year}
  {2022}),\ 10.1002/adfm.202208974}\BibitemShut {NoStop}%
\bibitem [{\citenamefont {Kresse}\ and\ \citenamefont
  {Furthm{\"u}ller}(1996)}]{VASP}%
  \BibitemOpen
  \bibfield  {author} {\bibinfo {author} {\bibfnamefont {G.}~\bibnamefont
  {Kresse}}\ and\ \bibinfo {author} {\bibfnamefont {J.}~\bibnamefont
  {Furthm{\"u}ller}},\ }\href@noop {} {\bibfield  {journal} {\bibinfo
  {journal} {Physical Review B}\ }\textbf {\bibinfo {volume} {54}},\ \bibinfo
  {pages} {11169} (\bibinfo {year} {1996})}\BibitemShut {NoStop}%
\bibitem [{\citenamefont {Perdew}\ \emph {et~al.}(1996)\citenamefont {Perdew},
  \citenamefont {Burke},\ and\ \citenamefont
  {Ernzerhof}}]{perdew1996generalized}%
  \BibitemOpen
  \bibfield  {author} {\bibinfo {author} {\bibfnamefont {J.~P.}\ \bibnamefont
  {Perdew}}, \bibinfo {author} {\bibfnamefont {K.}~\bibnamefont {Burke}}, \
  and\ \bibinfo {author} {\bibfnamefont {M.}~\bibnamefont {Ernzerhof}},\
  }\href@noop {} {\bibfield  {journal} {\bibinfo  {journal} {Physical Review
  Letters}\ }\textbf {\bibinfo {volume} {77}},\ \bibinfo {pages} {3865}
  (\bibinfo {year} {1996})}\BibitemShut {NoStop}%
\bibitem [{\citenamefont {Cuono}\ \emph {et~al.}(2023)\citenamefont {Cuono},
  \citenamefont {Sattigeri}, \citenamefont {Autieri},\ and\ \citenamefont
  {Dietl}}]{Cuono23Eu}%
  \BibitemOpen
  \bibfield  {author} {\bibinfo {author} {\bibfnamefont {G.}~\bibnamefont
  {Cuono}}, \bibinfo {author} {\bibfnamefont {R.~M.}\ \bibnamefont
  {Sattigeri}}, \bibinfo {author} {\bibfnamefont {C.}~\bibnamefont {Autieri}},
  \ and\ \bibinfo {author} {\bibfnamefont {T.}~\bibnamefont {Dietl}},\ }\href
  {\doibase 10.1103/PhysRevB.108.075150} {\bibfield  {journal} {\bibinfo
  {journal} {Phys. Rev. B}\ }\textbf {\bibinfo {volume} {108}},\ \bibinfo
  {pages} {075150} (\bibinfo {year} {2023})}\BibitemShut {NoStop}%
\bibitem [{\citenamefont {Islam}\ \emph {et~al.}(2019)\citenamefont {Islam},
  \citenamefont {Cuono}, \citenamefont {Nguyen}, \citenamefont {Noce},\ and\
  \citenamefont {Autieri}}]{Islam19}%
  \BibitemOpen
  \bibfield  {author} {\bibinfo {author} {\bibfnamefont {R.}~\bibnamefont
  {Islam}}, \bibinfo {author} {\bibfnamefont {G.}~\bibnamefont {Cuono}},
  \bibinfo {author} {\bibfnamefont {M.~N.}\ \bibnamefont {Nguyen}}, \bibinfo
  {author} {\bibfnamefont {C.}~\bibnamefont {Noce}}, \ and\ \bibinfo {author}
  {\bibfnamefont {C.}~\bibnamefont {Autieri}},\ }\href {\doibase
  10.12693/APhysPolA.136.667} {\bibfield  {journal} {\bibinfo  {journal} {Acta
  Physica Polonica A}\ }\textbf {\bibinfo {volume} {136}} (\bibinfo {year}
  {2019}),\ 10.12693/APhysPolA.136.667}\BibitemShut {NoStop}%
\bibitem [{\citenamefont {Hussain}\ \emph {et~al.}(2022)\citenamefont
  {Hussain}, \citenamefont {Cuono}, \citenamefont {Islam}, \citenamefont
  {Trajnerowicz}, \citenamefont {Jureńczyk}, \citenamefont {Autieri},\ and\
  \citenamefont {Dietl}}]{Hussain22}%
  \BibitemOpen
  \bibfield  {author} {\bibinfo {author} {\bibfnamefont {G.}~\bibnamefont
  {Hussain}}, \bibinfo {author} {\bibfnamefont {G.}~\bibnamefont {Cuono}},
  \bibinfo {author} {\bibfnamefont {R.}~\bibnamefont {Islam}}, \bibinfo
  {author} {\bibfnamefont {A.}~\bibnamefont {Trajnerowicz}}, \bibinfo {author}
  {\bibfnamefont {J.}~\bibnamefont {Jureńczyk}}, \bibinfo {author}
  {\bibfnamefont {C.}~\bibnamefont {Autieri}}, \ and\ \bibinfo {author}
  {\bibfnamefont {T.}~\bibnamefont {Dietl}},\ }\href {\doibase
  10.1088/1361-6463/ac984d} {\bibfield  {journal} {\bibinfo  {journal} {Journal
  of Physics D: Applied Physics}\ }\textbf {\bibinfo {volume} {55}},\ \bibinfo
  {pages} {495301} (\bibinfo {year} {2022})}\BibitemShut {NoStop}%
\bibitem [{\citenamefont {Sun}\ \emph {et~al.}(2015)\citenamefont {Sun},
  \citenamefont {Ruzsinszky},\ and\ \citenamefont
  {Perdew}}]{PhysRevLett.115.036402}%
  \BibitemOpen
  \bibfield  {author} {\bibinfo {author} {\bibfnamefont {J.}~\bibnamefont
  {Sun}}, \bibinfo {author} {\bibfnamefont {A.}~\bibnamefont {Ruzsinszky}}, \
  and\ \bibinfo {author} {\bibfnamefont {J.~P.}\ \bibnamefont {Perdew}},\
  }\href {\doibase 10.1103/PhysRevLett.115.036402} {\bibfield  {journal}
  {\bibinfo  {journal} {Phys. Rev. Lett.}\ }\textbf {\bibinfo {volume} {115}},\
  \bibinfo {pages} {036402} (\bibinfo {year} {2015})}\BibitemShut {NoStop}%
\bibitem [{\citenamefont {Skylaris}\ \emph {et~al.}(2005)\citenamefont
  {Skylaris}, \citenamefont {Haynes}, \citenamefont {Mostofi},\ and\
  \citenamefont {Payne}}]{10.1063/1.1839852}%
  \BibitemOpen
  \bibfield  {author} {\bibinfo {author} {\bibfnamefont {C.-K.}\ \bibnamefont
  {Skylaris}}, \bibinfo {author} {\bibfnamefont {P.~D.}\ \bibnamefont
  {Haynes}}, \bibinfo {author} {\bibfnamefont {A.~A.}\ \bibnamefont {Mostofi}},
  \ and\ \bibinfo {author} {\bibfnamefont {M.~C.}\ \bibnamefont {Payne}},\
  }\href {\doibase 10.1063/1.1839852} {\bibfield  {journal} {\bibinfo
  {journal} {The Journal of Chemical Physics}\ }\textbf {\bibinfo {volume}
  {122}},\ \bibinfo {pages} {084119} (\bibinfo {year} {2005})},\ \Eprint
  {http://arxiv.org/abs/https://pubs.aip.org/aip/jcp/article-pdf/doi/10.1063/1.1839852/15362896/084119\_1\_online.pdf}
  {https://pubs.aip.org/aip/jcp/article-pdf/doi/10.1063/1.1839852/15362896/084119\_1\_online.pdf}
  \BibitemShut {NoStop}%
\bibitem [{\citenamefont {Prentice}\ \emph {et~al.}(2020)\citenamefont
  {Prentice}, \citenamefont {Aarons}, \citenamefont {Womack}, \citenamefont
  {Allen}, \citenamefont {Andrinopoulos}, \citenamefont {Anton}, \citenamefont
  {Bell}, \citenamefont {Bhandari}, \citenamefont {Bramley}, \citenamefont
  {Charlton}, \citenamefont {Clements}, \citenamefont {Cole}, \citenamefont
  {Constantinescu}, \citenamefont {Corsetti}, \citenamefont {Dubois},
  \citenamefont {Duff}, \citenamefont {Escart\'in}, \citenamefont {Greco},
  \citenamefont {Hill}, \citenamefont {Lee}, \citenamefont {Linscott},
  \citenamefont {O’Regan}, \citenamefont {Phipps}, \citenamefont {Ratcliff},
  \citenamefont {Serrano}, \citenamefont {Tait}, \citenamefont {Teobaldi},
  \citenamefont {Vitale}, \citenamefont {Yeung}, \citenamefont {Zuehlsdorff},
  \citenamefont {Dziedzic}, \citenamefont {Haynes}, \citenamefont {Hine},
  \citenamefont {Mostofi}, \citenamefont {Payne},\ and\ \citenamefont
  {Skylaris}}]{10.1063/5.0004445}%
  \BibitemOpen
  \bibfield  {author} {\bibinfo {author} {\bibfnamefont {J.~C.~A.}\
  \bibnamefont {Prentice}}, \bibinfo {author} {\bibfnamefont {J.}~\bibnamefont
  {Aarons}}, \bibinfo {author} {\bibfnamefont {J.~C.}\ \bibnamefont {Womack}},
  \bibinfo {author} {\bibfnamefont {A.~E.~A.}\ \bibnamefont {Allen}}, \bibinfo
  {author} {\bibfnamefont {L.}~\bibnamefont {Andrinopoulos}}, \bibinfo {author}
  {\bibfnamefont {L.}~\bibnamefont {Anton}}, \bibinfo {author} {\bibfnamefont
  {R.~A.}\ \bibnamefont {Bell}}, \bibinfo {author} {\bibfnamefont
  {A.}~\bibnamefont {Bhandari}}, \bibinfo {author} {\bibfnamefont {G.~A.}\
  \bibnamefont {Bramley}}, \bibinfo {author} {\bibfnamefont {R.~J.}\
  \bibnamefont {Charlton}}, \bibinfo {author} {\bibfnamefont {R.~J.}\
  \bibnamefont {Clements}}, \bibinfo {author} {\bibfnamefont {D.~J.}\
  \bibnamefont {Cole}}, \bibinfo {author} {\bibfnamefont {G.}~\bibnamefont
  {Constantinescu}}, \bibinfo {author} {\bibfnamefont {F.}~\bibnamefont
  {Corsetti}}, \bibinfo {author} {\bibfnamefont {S.~M.-M.}\ \bibnamefont
  {Dubois}}, \bibinfo {author} {\bibfnamefont {K.~K.~B.}\ \bibnamefont {Duff}},
  \bibinfo {author} {\bibfnamefont {J.~M.}\ \bibnamefont {Escart\'in}},
  \bibinfo {author} {\bibfnamefont {A.}~\bibnamefont {Greco}}, \bibinfo
  {author} {\bibfnamefont {Q.}~\bibnamefont {Hill}}, \bibinfo {author}
  {\bibfnamefont {L.~P.}\ \bibnamefont {Lee}}, \bibinfo {author} {\bibfnamefont
  {E.}~\bibnamefont {Linscott}}, \bibinfo {author} {\bibfnamefont {D.~D.}\
  \bibnamefont {O’Regan}}, \bibinfo {author} {\bibfnamefont {M.~J.~S.}\
  \bibnamefont {Phipps}}, \bibinfo {author} {\bibfnamefont {L.~E.}\
  \bibnamefont {Ratcliff}}, \bibinfo {author} {\bibfnamefont {A.~R.}\
  \bibnamefont {Serrano}}, \bibinfo {author} {\bibfnamefont {E.~W.}\
  \bibnamefont {Tait}}, \bibinfo {author} {\bibfnamefont {G.}~\bibnamefont
  {Teobaldi}}, \bibinfo {author} {\bibfnamefont {V.}~\bibnamefont {Vitale}},
  \bibinfo {author} {\bibfnamefont {N.}~\bibnamefont {Yeung}}, \bibinfo
  {author} {\bibfnamefont {T.~J.}\ \bibnamefont {Zuehlsdorff}}, \bibinfo
  {author} {\bibfnamefont {J.}~\bibnamefont {Dziedzic}}, \bibinfo {author}
  {\bibfnamefont {P.~D.}\ \bibnamefont {Haynes}}, \bibinfo {author}
  {\bibfnamefont {N.~D.~M.}\ \bibnamefont {Hine}}, \bibinfo {author}
  {\bibfnamefont {A.~A.}\ \bibnamefont {Mostofi}}, \bibinfo {author}
  {\bibfnamefont {M.~C.}\ \bibnamefont {Payne}}, \ and\ \bibinfo {author}
  {\bibfnamefont {C.-K.}\ \bibnamefont {Skylaris}},\ }\href {\doibase
  10.1063/5.0004445} {\bibfield  {journal} {\bibinfo  {journal} {The Journal of
  Chemical Physics}\ }\textbf {\bibinfo {volume} {152}},\ \bibinfo {pages}
  {174111} (\bibinfo {year} {2020})},\ \Eprint
  {http://arxiv.org/abs/https://pubs.aip.org/aip/jcp/article-pdf/doi/10.1063/5.0004445/16741083/174111\_1\_online.pdf}
  {https://pubs.aip.org/aip/jcp/article-pdf/doi/10.1063/5.0004445/16741083/174111\_1\_online.pdf}
  \BibitemShut {NoStop}%
\bibitem [{\citenamefont {Skylaris}\ and\ \citenamefont
  {Haynes}(2007)}]{10.1063/1.2796168}%
  \BibitemOpen
  \bibfield  {author} {\bibinfo {author} {\bibfnamefont {C.-K.}\ \bibnamefont
  {Skylaris}}\ and\ \bibinfo {author} {\bibfnamefont {P.~D.}\ \bibnamefont
  {Haynes}},\ }\href {\doibase 10.1063/1.2796168} {\bibfield  {journal}
  {\bibinfo  {journal} {The Journal of Chemical Physics}\ }\textbf {\bibinfo
  {volume} {127}},\ \bibinfo {pages} {164712} (\bibinfo {year} {2007})},\
  \Eprint
  {http://arxiv.org/abs/https://pubs.aip.org/aip/jcp/article-pdf/doi/10.1063/1.2796168/15403114/164712\_1\_online.pdf}
  {https://pubs.aip.org/aip/jcp/article-pdf/doi/10.1063/1.2796168/15403114/164712\_1\_online.pdf}
  \BibitemShut {NoStop}%
\bibitem [{\citenamefont {Skylaris}\ \emph {et~al.}(2008)\citenamefont
  {Skylaris}, \citenamefont {Haynes}, \citenamefont {Mostofi},\ and\
  \citenamefont {Payne}}]{Skylaris_2008}%
  \BibitemOpen
  \bibfield  {author} {\bibinfo {author} {\bibfnamefont {C.-K.}\ \bibnamefont
  {Skylaris}}, \bibinfo {author} {\bibfnamefont {P.~D.}\ \bibnamefont
  {Haynes}}, \bibinfo {author} {\bibfnamefont {A.~A.}\ \bibnamefont {Mostofi}},
  \ and\ \bibinfo {author} {\bibfnamefont {M.~C.}\ \bibnamefont {Payne}},\
  }\href {\doibase 10.1088/0953-8984/20/6/064209} {\bibfield  {journal}
  {\bibinfo  {journal} {Journal of Physics: Condensed Matter}\ }\textbf
  {\bibinfo {volume} {20}},\ \bibinfo {pages} {064209} (\bibinfo {year}
  {2008})}\BibitemShut {NoStop}%
\bibitem [{\citenamefont {Jollet}\ \emph {et~al.}(2014)\citenamefont {Jollet},
  \citenamefont {Torrent},\ and\ \citenamefont {Holzwarth}}]{JOLLET20141246}%
  \BibitemOpen
  \bibfield  {author} {\bibinfo {author} {\bibfnamefont {F.}~\bibnamefont
  {Jollet}}, \bibinfo {author} {\bibfnamefont {M.}~\bibnamefont {Torrent}}, \
  and\ \bibinfo {author} {\bibfnamefont {N.}~\bibnamefont {Holzwarth}},\ }\href
  {\doibase https://doi.org/10.1016/j.cpc.2013.12.023} {\bibfield  {journal}
  {\bibinfo  {journal} {Computer Physics Communications}\ }\textbf {\bibinfo
  {volume} {185}},\ \bibinfo {pages} {1246} (\bibinfo {year}
  {2014})}\BibitemShut {NoStop}%
\bibitem [{\citenamefont {Constantinescu}\ and\ \citenamefont
  {Hine}(2015)}]{PhysRevB.91.195416}%
  \BibitemOpen
  \bibfield  {author} {\bibinfo {author} {\bibfnamefont {G.~C.}\ \bibnamefont
  {Constantinescu}}\ and\ \bibinfo {author} {\bibfnamefont {N.~D.~M.}\
  \bibnamefont {Hine}},\ }\href {\doibase 10.1103/PhysRevB.91.195416}
  {\bibfield  {journal} {\bibinfo  {journal} {Phys. Rev. B}\ }\textbf {\bibinfo
  {volume} {91}},\ \bibinfo {pages} {195416} (\bibinfo {year}
  {2015})}\BibitemShut {NoStop}%
\bibitem [{\citenamefont {Okamura}\ \emph {et~al.}(2022)\citenamefont
  {Okamura}, \citenamefont {Handa}, \citenamefont {Yoshimi}, \citenamefont
  {Tsukazaki}, \citenamefont {Takahashi}, \citenamefont {Kawasaki},
  \citenamefont {Tokura},\ and\ \citenamefont {Takahashi}}]{Okamura23}%
  \BibitemOpen
  \bibfield  {author} {\bibinfo {author} {\bibfnamefont {Y.}~\bibnamefont
  {Okamura}}, \bibinfo {author} {\bibfnamefont {H.}~\bibnamefont {Handa}},
  \bibinfo {author} {\bibfnamefont {R.}~\bibnamefont {Yoshimi}}, \bibinfo
  {author} {\bibfnamefont {A.}~\bibnamefont {Tsukazaki}}, \bibinfo {author}
  {\bibfnamefont {K.~S.}\ \bibnamefont {Takahashi}}, \bibinfo {author}
  {\bibfnamefont {M.}~\bibnamefont {Kawasaki}}, \bibinfo {author}
  {\bibfnamefont {Y.}~\bibnamefont {Tokura}}, \ and\ \bibinfo {author}
  {\bibfnamefont {Y.}~\bibnamefont {Takahashi}},\ }\href {\doibase
  10.1038/s41535-022-00501-2} {\bibfield  {journal} {\bibinfo  {journal} {npj
  Quantum Materials}\ }\textbf {\bibinfo {volume} {7}} (\bibinfo {year}
  {2022}),\ 10.1038/s41535-022-00501-2}\BibitemShut {NoStop}%
\bibitem [{\citenamefont {Tanaka}\ \emph {et~al.}(2013)\citenamefont {Tanaka},
  \citenamefont {Shoman}, \citenamefont {Nakayama}, \citenamefont {Souma},
  \citenamefont {Sato}, \citenamefont {Takahashi}, \citenamefont {Novak},
  \citenamefont {Segawa},\ and\ \citenamefont {Ando}}]{PhysRevB.88.235126}%
  \BibitemOpen
  \bibfield  {author} {\bibinfo {author} {\bibfnamefont {Y.}~\bibnamefont
  {Tanaka}}, \bibinfo {author} {\bibfnamefont {T.}~\bibnamefont {Shoman}},
  \bibinfo {author} {\bibfnamefont {K.}~\bibnamefont {Nakayama}}, \bibinfo
  {author} {\bibfnamefont {S.}~\bibnamefont {Souma}}, \bibinfo {author}
  {\bibfnamefont {T.}~\bibnamefont {Sato}}, \bibinfo {author} {\bibfnamefont
  {T.}~\bibnamefont {Takahashi}}, \bibinfo {author} {\bibfnamefont
  {M.}~\bibnamefont {Novak}}, \bibinfo {author} {\bibfnamefont
  {K.}~\bibnamefont {Segawa}}, \ and\ \bibinfo {author} {\bibfnamefont
  {Y.}~\bibnamefont {Ando}},\ }\href {\doibase 10.1103/PhysRevB.88.235126}
  {\bibfield  {journal} {\bibinfo  {journal} {Phys. Rev. B}\ }\textbf {\bibinfo
  {volume} {88}},\ \bibinfo {pages} {235126} (\bibinfo {year}
  {2013})}\BibitemShut {NoStop}%
\bibitem [{\citenamefont {Xu}\ \emph {et~al.}(2015)\citenamefont {Xu},
  \citenamefont {Li}, \citenamefont {Martinez}, \citenamefont {Sinitsyn},
  \citenamefont {Htoon}, \citenamefont {Li}, \citenamefont {Swartzentruber},
  \citenamefont {Hollingsworth}, \citenamefont {Wang},\ and\ \citenamefont
  {Zhang}}]{C4NR05870D}%
  \BibitemOpen
  \bibfield  {author} {\bibinfo {author} {\bibfnamefont {E.~Z.}\ \bibnamefont
  {Xu}}, \bibinfo {author} {\bibfnamefont {Z.}~\bibnamefont {Li}}, \bibinfo
  {author} {\bibfnamefont {J.~A.}\ \bibnamefont {Martinez}}, \bibinfo {author}
  {\bibfnamefont {N.}~\bibnamefont {Sinitsyn}}, \bibinfo {author}
  {\bibfnamefont {H.}~\bibnamefont {Htoon}}, \bibinfo {author} {\bibfnamefont
  {N.}~\bibnamefont {Li}}, \bibinfo {author} {\bibfnamefont {B.}~\bibnamefont
  {Swartzentruber}}, \bibinfo {author} {\bibfnamefont {J.~A.}\ \bibnamefont
  {Hollingsworth}}, \bibinfo {author} {\bibfnamefont {J.}~\bibnamefont {Wang}},
  \ and\ \bibinfo {author} {\bibfnamefont {S.~X.}\ \bibnamefont {Zhang}},\
  }\href {\doibase 10.1039/C4NR05870D} {\bibfield  {journal} {\bibinfo
  {journal} {Nanoscale}\ }\textbf {\bibinfo {volume} {7}},\ \bibinfo {pages}
  {2869} (\bibinfo {year} {2015})}\BibitemShut {NoStop}%
\bibitem [{\citenamefont {Skiff}\ \emph
  {et~al.}(2023{\natexlab{b}})\citenamefont {Skiff}, \citenamefont {de~Juan},
  \citenamefont {Queiroz}, \citenamefont {Mathimalar}, \citenamefont
  {Beidenkopf},\ and\ \citenamefont {Ilan}}]{10.21468/SciPostPhysCore.6.1.011}%
  \BibitemOpen
  \bibfield  {author} {\bibinfo {author} {\bibfnamefont {R.~M.}\ \bibnamefont
  {Skiff}}, \bibinfo {author} {\bibfnamefont {F.}~\bibnamefont {de~Juan}},
  \bibinfo {author} {\bibfnamefont {R.}~\bibnamefont {Queiroz}}, \bibinfo
  {author} {\bibfnamefont {S.}~\bibnamefont {Mathimalar}}, \bibinfo {author}
  {\bibfnamefont {H.}~\bibnamefont {Beidenkopf}}, \ and\ \bibinfo {author}
  {\bibfnamefont {R.}~\bibnamefont {Ilan}},\ }\href {\doibase
  10.21468/SciPostPhysCore.6.1.011} {\bibfield  {journal} {\bibinfo  {journal}
  {SciPost Phys. Core}\ }\textbf {\bibinfo {volume} {6}},\ \bibinfo {pages}
  {011} (\bibinfo {year} {2023}{\natexlab{b}})}\BibitemShut {NoStop}%
\bibitem [{\citenamefont {Si}\ \emph {et~al.}(2020)\citenamefont {Si},
  \citenamefont {Li}, \citenamefont {Zhou}, \citenamefont {Liu},\ and\
  \citenamefont {Prezhdo}}]{10.1063/1.5134951}%
  \BibitemOpen
  \bibfield  {author} {\bibinfo {author} {\bibfnamefont {Y.}~\bibnamefont
  {Si}}, \bibinfo {author} {\bibfnamefont {M.}~\bibnamefont {Li}}, \bibinfo
  {author} {\bibfnamefont {Z.}~\bibnamefont {Zhou}}, \bibinfo {author}
  {\bibfnamefont {M.}~\bibnamefont {Liu}}, \ and\ \bibinfo {author}
  {\bibfnamefont {O.}~\bibnamefont {Prezhdo}},\ }\href {\doibase
  10.1063/1.5134951} {\bibfield  {journal} {\bibinfo  {journal} {The Journal of
  Chemical Physics}\ }\textbf {\bibinfo {volume} {152}},\ \bibinfo {pages}
  {024706} (\bibinfo {year} {2020})},\ \Eprint
  {http://arxiv.org/abs/https://pubs.aip.org/aip/jcp/article-pdf/doi/10.1063/1.5134951/15572775/024706\_1\_online.pdf}
  {https://pubs.aip.org/aip/jcp/article-pdf/doi/10.1063/1.5134951/15572775/024706\_1\_online.pdf}
  \BibitemShut {NoStop}%
\bibitem [{\citenamefont {Hoskam}(2018)}]{Delft_thesis}%
  \BibitemOpen
  \bibfield  {author} {\bibinfo {author} {\bibfnamefont {M.~S.~M.}\
  \bibnamefont {Hoskam}},\ }\href@noop {} {\enquote {\bibinfo {title} {{Theory
  and fabrication of SnTe for Majorana devices}},}\ } (\bibinfo {year}
  {2018})\BibitemShut {NoStop}%
\bibitem [{\citenamefont {Klett}\ \emph {et~al.}(2018)\citenamefont {Klett},
  \citenamefont {Schönle}, \citenamefont {Becker}, \citenamefont {Dyck},
  \citenamefont {Borisov}, \citenamefont {Rott}, \citenamefont {Ramermann},
  \citenamefont {Büker}, \citenamefont {Haskenhoff}, \citenamefont {Krieft},
  \citenamefont {Hübner}, \citenamefont {Reimer}, \citenamefont {Shekhar},
  \citenamefont {Schmalhorst}, \citenamefont {Hütten}, \citenamefont {Felser},
  \citenamefont {Wernsdorfer},\ and\ \citenamefont {Reiss}}]{Klett18}%
  \BibitemOpen
  \bibfield  {author} {\bibinfo {author} {\bibfnamefont {R.}~\bibnamefont
  {Klett}}, \bibinfo {author} {\bibfnamefont {J.}~\bibnamefont {Schönle}},
  \bibinfo {author} {\bibfnamefont {A.}~\bibnamefont {Becker}}, \bibinfo
  {author} {\bibfnamefont {D.}~\bibnamefont {Dyck}}, \bibinfo {author}
  {\bibfnamefont {K.}~\bibnamefont {Borisov}}, \bibinfo {author} {\bibfnamefont
  {K.}~\bibnamefont {Rott}}, \bibinfo {author} {\bibfnamefont {D.}~\bibnamefont
  {Ramermann}}, \bibinfo {author} {\bibfnamefont {B.}~\bibnamefont {Büker}},
  \bibinfo {author} {\bibfnamefont {J.}~\bibnamefont {Haskenhoff}}, \bibinfo
  {author} {\bibfnamefont {J.}~\bibnamefont {Krieft}}, \bibinfo {author}
  {\bibfnamefont {T.}~\bibnamefont {Hübner}}, \bibinfo {author} {\bibfnamefont
  {O.}~\bibnamefont {Reimer}}, \bibinfo {author} {\bibfnamefont
  {C.}~\bibnamefont {Shekhar}}, \bibinfo {author} {\bibfnamefont {J.-M.}\
  \bibnamefont {Schmalhorst}}, \bibinfo {author} {\bibfnamefont
  {A.}~\bibnamefont {Hütten}}, \bibinfo {author} {\bibfnamefont
  {C.}~\bibnamefont {Felser}}, \bibinfo {author} {\bibfnamefont
  {W.}~\bibnamefont {Wernsdorfer}}, \ and\ \bibinfo {author} {\bibfnamefont
  {G.}~\bibnamefont {Reiss}},\ }\href {\doibase 10.1021/acs.nanolett.7b04870}
  {\bibfield  {journal} {\bibinfo  {journal} {Nano Letters}\ }\textbf {\bibinfo
  {volume} {18}},\ \bibinfo {pages} {1264} (\bibinfo {year}
  {2018})}\BibitemShut {NoStop}%
\bibitem [{\citenamefont {Mitobe}\ \emph {et~al.}(2021)\citenamefont {Mitobe},
  \citenamefont {Hoshi}, \citenamefont {Kasem}, \citenamefont {Kiyama},
  \citenamefont {Usui}, \citenamefont {Yamashita}, \citenamefont {Higashinaka},
  \citenamefont {Matsuda}, \citenamefont {Aoki}, \citenamefont {Katase},
  \citenamefont {Goto},\ and\ \citenamefont {Mizuguchi}}]{Mitobe21}%
  \BibitemOpen
  \bibfield  {author} {\bibinfo {author} {\bibfnamefont {T.}~\bibnamefont
  {Mitobe}}, \bibinfo {author} {\bibfnamefont {K.}~\bibnamefont {Hoshi}},
  \bibinfo {author} {\bibfnamefont {M.~R.}\ \bibnamefont {Kasem}}, \bibinfo
  {author} {\bibfnamefont {R.}~\bibnamefont {Kiyama}}, \bibinfo {author}
  {\bibfnamefont {H.}~\bibnamefont {Usui}}, \bibinfo {author} {\bibfnamefont
  {A.}~\bibnamefont {Yamashita}}, \bibinfo {author} {\bibfnamefont
  {R.}~\bibnamefont {Higashinaka}}, \bibinfo {author} {\bibfnamefont {T.~D.}\
  \bibnamefont {Matsuda}}, \bibinfo {author} {\bibfnamefont {Y.}~\bibnamefont
  {Aoki}}, \bibinfo {author} {\bibfnamefont {T.}~\bibnamefont {Katase}},
  \bibinfo {author} {\bibfnamefont {Y.}~\bibnamefont {Goto}}, \ and\ \bibinfo
  {author} {\bibfnamefont {Y.}~\bibnamefont {Mizuguchi}},\ }\href {\doibase
  10.1038/s41598-021-02341-9} {\bibfield  {journal} {\bibinfo  {journal}
  {Scientific Reports}\ }\textbf {\bibinfo {volume} {11}} (\bibinfo {year}
  {2021}),\ 10.1038/s41598-021-02341-9}\BibitemShut {NoStop}%
\bibitem [{\citenamefont {Fogel}\ \emph {et~al.}(2001)\citenamefont {Fogel},
  \citenamefont {Pokhila}, \citenamefont {Bomze}, \citenamefont {Sipatov},
  \citenamefont {Fedorenko},\ and\ \citenamefont {Shekhter}}]{Fogel01}%
  \BibitemOpen
  \bibfield  {author} {\bibinfo {author} {\bibfnamefont {N.~Y.}\ \bibnamefont
  {Fogel}}, \bibinfo {author} {\bibfnamefont {A.~S.}\ \bibnamefont {Pokhila}},
  \bibinfo {author} {\bibfnamefont {Y.~V.}\ \bibnamefont {Bomze}}, \bibinfo
  {author} {\bibfnamefont {A.~Y.}\ \bibnamefont {Sipatov}}, \bibinfo {author}
  {\bibfnamefont {A.~I.}\ \bibnamefont {Fedorenko}}, \ and\ \bibinfo {author}
  {\bibfnamefont {R.~I.}\ \bibnamefont {Shekhter}},\ }\href {\doibase
  10.1103/PhysRevLett.86.512} {\bibfield  {journal} {\bibinfo  {journal} {Phys.
  Rev. Lett.}\ }\textbf {\bibinfo {volume} {86}},\ \bibinfo {pages} {512}
  (\bibinfo {year} {2001})}\BibitemShut {NoStop}%
\bibitem [{\citenamefont {Tang}\ and\ \citenamefont {Fu}(2014)}]{Tang14}%
  \BibitemOpen
  \bibfield  {author} {\bibinfo {author} {\bibfnamefont {E.}~\bibnamefont
  {Tang}}\ and\ \bibinfo {author} {\bibfnamefont {L.}~\bibnamefont {Fu}},\
  }\href {\doibase 10.1038/nphys3109} {\bibfield  {journal} {\bibinfo
  {journal} {Nature Physics}\ }\textbf {\bibinfo {volume} {10}} (\bibinfo
  {year} {2014}),\ 10.1038/nphys3109}\BibitemShut {NoStop}%
\bibitem [{\citenamefont {Wang}\ \emph {et~al.}(2023)\citenamefont {Wang},
  \citenamefont {Li}, \citenamefont {Tian}, \citenamefont {Sang}, \citenamefont
  {Zhou}, \citenamefont {Chen}, \citenamefont {Zhao}, \citenamefont {Wu},
  \citenamefont {Zhang}, \citenamefont {Bellaiche}, \citenamefont {Liu},\ and\
  \citenamefont {Yang}}]{PhysRevB.108.045114}%
  \BibitemOpen
  \bibfield  {author} {\bibinfo {author} {\bibfnamefont {X.}~\bibnamefont
  {Wang}}, \bibinfo {author} {\bibfnamefont {X.}~\bibnamefont {Li}}, \bibinfo
  {author} {\bibfnamefont {H.}~\bibnamefont {Tian}}, \bibinfo {author}
  {\bibfnamefont {H.}~\bibnamefont {Sang}}, \bibinfo {author} {\bibfnamefont
  {J.}~\bibnamefont {Zhou}}, \bibinfo {author} {\bibfnamefont {L.}~\bibnamefont
  {Chen}}, \bibinfo {author} {\bibfnamefont {H.~J.}\ \bibnamefont {Zhao}},
  \bibinfo {author} {\bibfnamefont {D.}~\bibnamefont {Wu}}, \bibinfo {author}
  {\bibfnamefont {H.}~\bibnamefont {Zhang}}, \bibinfo {author} {\bibfnamefont
  {L.}~\bibnamefont {Bellaiche}}, \bibinfo {author} {\bibfnamefont {J.-M.}\
  \bibnamefont {Liu}}, \ and\ \bibinfo {author} {\bibfnamefont
  {Y.}~\bibnamefont {Yang}},\ }\href {\doibase 10.1103/PhysRevB.108.045114}
  {\bibfield  {journal} {\bibinfo  {journal} {Phys. Rev. B}\ }\textbf {\bibinfo
  {volume} {108}},\ \bibinfo {pages} {045114} (\bibinfo {year}
  {2023})}\BibitemShut {NoStop}%
\bibitem [{\citenamefont {Shimada}\ \emph {et~al.}(2020)\citenamefont
  {Shimada}, \citenamefont {Minaguro}, \citenamefont {Xu}, \citenamefont
  {Wang},\ and\ \citenamefont {Kitamura}}]{nano10040732}%
  \BibitemOpen
  \bibfield  {author} {\bibinfo {author} {\bibfnamefont {T.}~\bibnamefont
  {Shimada}}, \bibinfo {author} {\bibfnamefont {K.}~\bibnamefont {Minaguro}},
  \bibinfo {author} {\bibfnamefont {T.}~\bibnamefont {Xu}}, \bibinfo {author}
  {\bibfnamefont {J.}~\bibnamefont {Wang}}, \ and\ \bibinfo {author}
  {\bibfnamefont {T.}~\bibnamefont {Kitamura}},\ }\href {\doibase
  10.3390/nano10040732} {\bibfield  {journal} {\bibinfo  {journal}
  {Nanomaterials}\ }\textbf {\bibinfo {volume} {10}} (\bibinfo {year} {2020}),\
  10.3390/nano10040732}\BibitemShut {NoStop}%
\bibitem [{\citenamefont {Safdar}\ \emph {et~al.}(2013)\citenamefont {Safdar},
  \citenamefont {Wang}, \citenamefont {Mirza}, \citenamefont {Wang},
  \citenamefont {Xu},\ and\ \citenamefont {He}}]{safdar2013topological}%
  \BibitemOpen
  \bibfield  {author} {\bibinfo {author} {\bibfnamefont {M.}~\bibnamefont
  {Safdar}}, \bibinfo {author} {\bibfnamefont {Q.}~\bibnamefont {Wang}},
  \bibinfo {author} {\bibfnamefont {M.}~\bibnamefont {Mirza}}, \bibinfo
  {author} {\bibfnamefont {Z.}~\bibnamefont {Wang}}, \bibinfo {author}
  {\bibfnamefont {K.}~\bibnamefont {Xu}}, \ and\ \bibinfo {author}
  {\bibfnamefont {J.}~\bibnamefont {He}},\ }\href@noop {} {\bibfield  {journal}
  {\bibinfo  {journal} {Nano letters}\ }\textbf {\bibinfo {volume} {13}},\
  \bibinfo {pages} {5344} (\bibinfo {year} {2013})}\BibitemShut {NoStop}%
\end{thebibliography}%

\end{document}